\documentclass[prb,aps,epsfig,showpacs,superscriptaddress,twocoloumn,15point,reprint]{revtex4-2}
\usepackage{graphicx}
\usepackage{dcolumn}
\usepackage{bm}
\makeatletter
\usepackage{amsmath}
\usepackage{epstopdf}
\usepackage{lineno,hyperref}
\usepackage{array}
\usepackage{xcolor}

\newcommand{\Rmnum}[1]{\expandafter\@slowromancap\romannumeral #1@}
\makeatother
\begin{document}
\title{\bf Unconventional anomalous Hall effect in hexagonal polar magnet $\mathrm{Y}_3\mathrm{Co}_8\mathrm{Sn}_4$}
\author{Afsar Ahmed  }
\affiliation{CMP Division, Saha Institute of Nuclear Physics, A CI of Homi Bhabha National Institute, Kolkata 700064, India}
\author{Jyoti Sharma}
\affiliation{Department of Physics, Indian Institute of Technology Bombay, Mumbai 400076, India}
\author{Arnab Bhattacharya }
\affiliation{CMP Division, Saha Institute of Nuclear Physics, A CI of Homi Bhabha National Institute, Kolkata 700064, India}
\author{Anis Biswas}
\affiliation{Ames National Laboratory, Iowa State University, Ames, Iowa 50011, USA}
\author{Tukai Singha}
\affiliation{CMP Division, Saha Institute of Nuclear Physics, A CI of Homi Bhabha National Institute, Kolkata 700064, India}
\author{Yaroslav Mudryk}
\affiliation{Ames National Laboratory, Iowa State University, Ames, Iowa 50011, USA}
\author{Aftab Alam}\email{aftab@iitb.ac.in}
\affiliation{Department of Physics, Indian Institute of Technology Bombay, Mumbai 400076, India}
\author{ I. Das}\email{indranil.das@saha.ac.in}
\affiliation{CMP Division, Saha Institute of Nuclear Physics, A CI of Homi Bhabha National Institute, Kolkata 700064, India}

\begin{abstract}

We report a rare realization of unconventional anomalous Hall effect (UAHE) both below and above the magnetic transition temperature ($T_C$) in a hexagonal noncentrosymmetric magnet $\mathrm{Y}_3\mathrm{Co}_8\mathrm{Sn}_4$,  using a combined
experimental and {\it ab-initio} calculations. Occurrence of such UAHE is mainly attributed to the reciprocal ($\mathcal{KS}$) topology (i.e. the presence of topological Weyl points at/near the Fermi level), along with some contribution from the topological magnetic texture, as inferred from the measured field-dependent ac susceptibility. The  effect of UAHE on the measured transport behavior however evolves differently with temperature above and below $T_C$, suggesting different physical mechanism responsible in the two phases. A unique planar ferrimagnetic ordering is found to be the most stable state with $ab$-plane as the easy plane below $T_C$, as observed experimentally. The simulated net magnetization and the moment per Co atom agrees fairly well with the measured values. A reasonably
large AHC is also observed in both the phases (above and below and $T_C$) of the present compound, which is again not so ubiquitous. Our results underscore the family of $\mathrm{R}_3\mathrm{Co}_8\mathrm{Sn}_4$ ($\mathrm{R}$= rare earth) polar magnets as a compelling backdrop for exploring the synergy of topological magnetism and non-trivial electronic bands, pivotal for spintronic applications.


\end{abstract}

\maketitle

{\it \bf Introduction:}
Integrating topological concepts with magnetism has unveiled a unique perspective for exploring emergent phenomena in quantum materials \cite{science.aav2873,keimer2017physics,science.aav2873, belopolski2019discovery,wang2018large,hasan2010colloquium}. Apart from the conventional anomalous Hall effect (AHE), sometimes an additional unconventional component have observed in these topological quantum materials, named as unconventional AHE (UAHE)\cite{neubauer2009topological}. In real space ($\mathcal{RS}$), UAHE has been reported in systems with non-coplanar spin arrangements characterized by finite topological number $\mathcal{Q} = \frac{1}{4\pi}\int \textbf{m}\cdot \left(\frac{\partial \textbf{m}}{\partial x}\times \frac{\partial \textbf{m}}{\partial y} \right) \textit{dxdy}$, and non-zero scaler spin chirality $\chi_{ijk}$ = \textbf{S$_i$}$\cdot$ (\textbf{S$_j$}$\times$\textbf{S$_k$}), where \textbf{S$_n$} and \textbf{m} are spins and unit vector of magnetization, respectively \cite{fert2017magnetic,heinze2011spontaneous}. Alongside, UAHE has also been observed in systems featuring topologically-secured band crossings, namely Weyl nodes, which always appear in pairs with opposite chirality \cite{shekhar2018anomalous,burkov2014anomalous,cheng2024tunable,xu2021unconventional,PhysRevLett.129.236601,PhysRevLett.124.017202,ybptbi2018}. These Weyl nodes located in the vicinity of the Fermi level, act as source and sink of Berry curvature while generating an effective magnetic field in reciprocal space ($\mathcal{KS}$). Weyl fermions arise either due to the broken inversion symmetry ($\mathcal{P}$) \cite{soluyanov2015type,yang2015weyl,lv2015experimental} or time-reversal symmetry ($\mathcal{T}$) \cite{xu2011chern,armitage2018weyl,wang2016time,science.aav2873}, providing shared prerequisite grounds for realizing UAHE\cite{tokura2020magnetic}. Generally, finding Weyl metallic state in a system with both $\mathcal{T}$ and $\mathcal{P}$ symmetry broken, is exotic and uncommon\cite{GdPtBinature, bhattacharya2024giant,chang2018magnetic}. This facilitates noncentrosymmetric magnets as a robust platform to study the interplay of topological magnetism and electronic phase, elicited through robust electrical transport responses.
At this juncture, $C_{nv}$ polar magnets owing to their broken $\mathcal{P}$, have gained tremendous attention both from theoretical\cite{chang2018magnetic} as well as experimental perspective \cite{li2023emergence}, to explore the interplay of $\mathcal{RS}$ and $\mathcal{KS}$ topology\cite{PhysRevLett.124.017202}.  
 
In this letter, we report the interplay between magnetism and topology in the non-centrosymmetric polar magnet $\mathrm{Y}_3\mathrm{Co}_8\mathrm{Sn}_4$ (\textit{C}$_{6v}$ symmetry) using a combined experimental and \textit{ab-initio} calculations. We observe an unconventional dual behavior of AHE, transiting from positive (below $T_C$) to negative (above $T_C$) sign. This reveals an unusual contribution either from non-trivial $\mathcal{KS}$ or $\mathcal{RS}$ topology or possibly a mixture of the two. Such behavior is rarely observed, specially hosting in the same system, and is in complete contrast to the conventional trend of anomalous transverse transport found in literature.
Below the magnetic transition temperature ($T_C$), our \textit{ab-initio} calculations confirm a unique ferrimagnetic (FiM) state hosting five pairs of Weyl nodes at/near the Fermi level (E$_F$), resulting in a large AHC value mediated by the intrinsic Berry curvature, which agrees fairly well with our experimental findings. We also simulated the high temperature  nonmagnetic (NM) phase of $\mathrm{Y}_3\mathrm{Co}_8\mathrm{Sn}_4$. Interestingly,  
four pairs of Weyl nodes are found at/near the Fermi level (E$_F$) 
 in this phase as well, which may be attributed to the breaking of the inversion symmetry. The effect of UAHE on the measured
transport behavior however remains quite different above and below $T_C$ , suggesting different physical mechanism responsible in the two phases. The simulated net magnetization and the moment per Co atom agrees fairly well with the
measured values. A reasonably large AHC is observed in both the phases (above and below and
$T_C$), which is not so ubiquitous. These findings underscore the emergence of novel functionalities driven by the coexistence of topological electronic bands and non-coplanar magnetic ordering.

{\it \bf Experimental and Computational Details:}
Polycrystalline  samples of $\mathrm{Y}_3\mathrm{Co}_8\mathrm{Sn}_4$ were prepared through the arc melting technique in a high-purity argon environment, employing stoichiometric quantities of high-purity (at least 99.9\%) constituent elements, followed by annealing at 800 $^{o}$C for four weeks in evacuated sealed quartz tube. First-principles calculations were performed using  Vienna \textit{Ab-initio} Simulation Package (VASP)\cite{PhysRev.136.B864} based on Density Functional Theory (DFT), with  Projected Augmented Wave (PAW)\cite{kresse1999ultrasoft} pseudopotential, and Perdew, Burke, and Ernzerhof (PBE) \cite{perdew1996generalized}functional. Further details about the sample preparation, experimental procedures, and computational methods are provided in the supplementary material (SM)\cite{SM}.

\begin{figure}[t]
\begin{center}
\includegraphics[width=.485\textwidth]{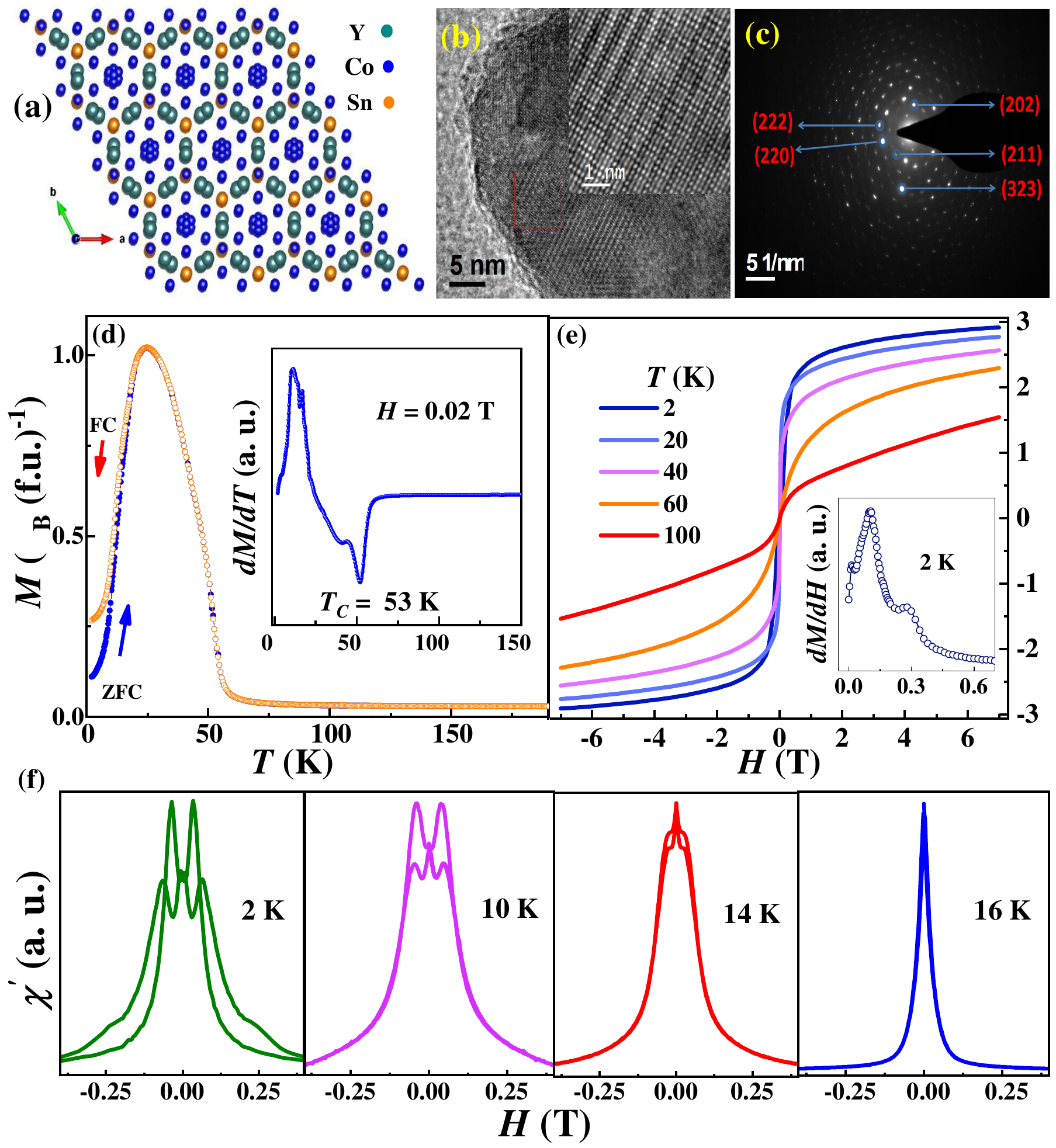}
\end{center}
\caption[]
{\label{F1} For $\mathrm{Y}_3\mathrm{Co}_8\mathrm{Sn}_4$, (a) Crystal structure (2D projection) (b)  HR-TEM image. Inset shows Fourier filtered image of marked region. (c) Selected-area electron diffraction (SAED) pattern. (d) Thermomagnetic $M(T)$ curve measured under 0.02 T. Inset shows d$M$/d$T$ vs. $T$ of ZFC. (e) Isothermal $M(H)$ at various $T$. Inset shows d$M$/d$H$ vs. $H$ for 2 K. (f) Real part of ac susceptibility ($\chi'$) vs. H for 2, 10, 12 and 16 K. }
\end{figure}

{\it \bf Results and Discussions:}
$\mathrm{Y}_3\mathrm{Co}_8\mathrm{Sn}_4$ crystallizes in polar hexagonal structure (space group: $P6_3mc$) as shown in Fig. \ref{F1}(a) (2D projection), a distorted derivative of centrosymmetric BaLi$_4$ structure (space group: $P6_3/mmc$). Lattice parameters, shown in Table ST1 of SM\cite{SM}, consistent with earlier reports\cite{canepa2000crystal,canepa2004ferromagnetic}. The High-resolution TEM (HRTEM) image and SAED pattern shown in Fig. \ref{F1}(b) and (c), respectively, reveal good crystalline nature of the samples. Thermomagnetic $M(T)$ measurements reveal a long-range ferromagnetic (FM) like transition at $T_C$ = 53 K, followed by a notable decrease and irreversibility between zero-field cooled (ZFC) and field-cooled (FC) curves below $T_K \sim$ 25 K, which is suppressed  at higher $H$ (see Fig.S2(a) of SM \cite{SM}). This resonates with the development of incommensurate antiferromagnetic (AFM) ordering at lower $T$\cite{canepa2000magnetic,canepa2004ferromagnetic,takagi2018multiple} (more details in SM\cite{SM}). The isothermal magnetization $M(H)$ attains a saturation moment of 2.54 $\mu_B/ f.u.$ (0.32 $\mu_\mathrm{B}$/Co atom) at $T$ = 2 K with a minor hysteresis(\cite{SM}), indicating the development of weak FM interaction~\cite{takagi2018multiple} (Fig. \ref{F1}(e)). Interestingly, $dM$/$dH$ at $T$ = 2 K reveals two distinct changes of slope, one at 60 mT and the other at 0.3 T, suggesting field-induced spin reorientations stemmed from competing AFM and FM interactions (more details in SM\cite{SM}). A peak/dip behaviour is reflected in the field-evolution of isothermal ac susceptibility $\chi'(H)$, as illustrated in Fig. \ref{F1}(f). From $\chi'(H)$ curves, a sharp peak at 0.1 T, followed by a hump-like behaviour at 0.3 T for 2 K is clearly evident. The peak positions gradually shift to lower field values with increasing T and vanish completely at 16 K. Notably, similar features are also observed across skyrmion/anti-skyrmion phase pockets in chiral cubic MnSi, centrosymmetric Gd$_2$PdSi$_3$ and several $D_{2d}$ Heusler alloys\cite{madduri2020ac,sen2019observation,p31,p30,p29,p28,madduri2020ac,p34}, suggesting the stabilization of topological magnetic phase in $\mathrm{Y}_3\mathrm{Co}_8\mathrm{Sn}_4$.

\begin{figure*}[t]
\begin{center}
\includegraphics[width=1\textwidth]{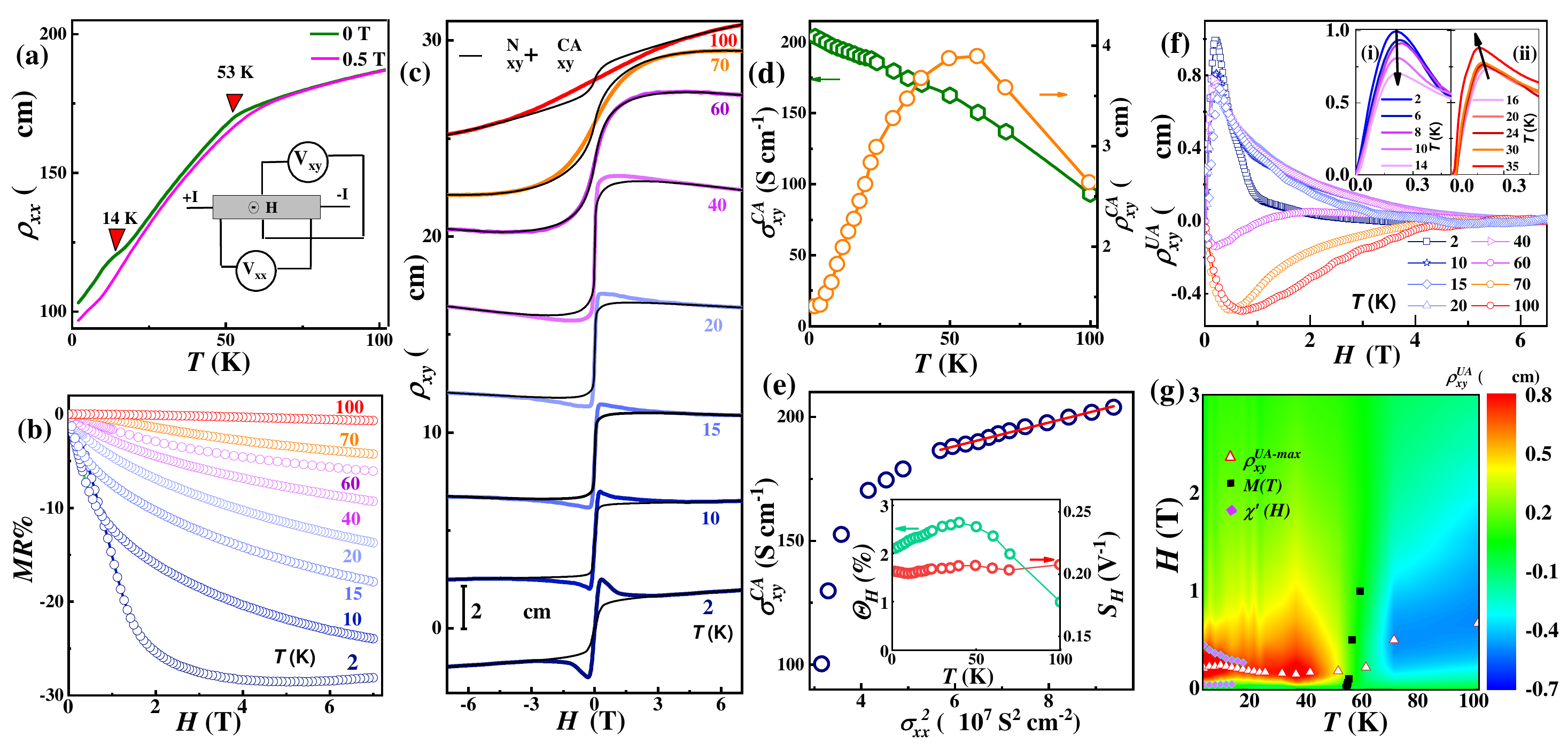}
\end{center}
\caption[]
{\label{F2}For $\mathrm{Y}_3\mathrm{Co}_8\mathrm{Sn}_4$, (a) $\rho_{xx}$ vs. $T$ under zero and 0.5 T applied field. Inset represent schematic of the orientation of sample with $H$. (b) Magnetoresistive (MR) isotherms at various $T$. (c) Total $\rho_{xy}$ vs. $H$ along with with the calculated curve including ordinary and conventional anomalous Hall components. A vertical offset is created for better visibility. (d) Conventional AHC (left axis) and AHR (right axis) vs. $T$. (e) $TYJ$ scaling of conventional AHC. Inset shows $T$ variation of $S_H$ and $\theta_H$. (f) Unconventional Hall resistivity ($\rho_{xy}^{UA}$) below and above $T_C$. Insets (i) and (ii) show an enlarged view of $\rho_{xy}^{UA}$ in the $T$ ranges 2 to 14 K and 16 to 35 K, respectively. (g) Contour plot of $\rho_{xy}^{UA}$ in $H$-$T$ phase diagram. }
\end{figure*}

We conducted systematic electrical transport measurements to elucidate the correlation between magnetism and the topological properties in both $\mathcal{RS}$ and $\mathcal{KS}$. Figure \ref{F2}(a) shows the $T$-dependence of longitudinal resistivity, $\rho_{xx}(T)$, exhibiting a kink at $T_C$ followed by a second anomaly at 14 K. The slope change in $\rho_{xx}$ near 14 K reflects a superzone gap effect from the incommensurate magnetic structure \cite{rawat2001magnetic,mishra2024evidence}, suppressed by an external magnetic field. Figure \ref{F2}(b) presents the field dependence of magnetoresistance (MR$\%$ = $\Delta\rho_{xx}(H)/\rho_{xx}(0)\times$ 100), revealing a pronounced negative saturating MR of approximately $\sim 30\%$ at 2 K. As $T$ increases, the MR decreases and transitions to non-saturating behaviour, reflecting the evolution from an incommensurate AFM phase to a field-induced FM like phase\cite{zhang2020anomalous,saha1999magnetic}.

Figure \ref{F2}(c) illustrates the field variation of transverse resistivity $\rho_{xy}$ isotherms. Intriguingly, a distinct anomaly appears in $\rho_{xy}$, superimposed on the background of normal ($\rho_{xy}^\mathrm{N}$) and conventional anomalous Hall resistivity ($\rho_{xy}^\mathrm{CA}$), a key feature associated with magnetic topological phase\cite{xu2021unconventional,p34,p35,neubauer2009topological}. For each isotherm, the empirical relation, $\rho_{xy} = \rho_{xy}^\mathrm{N} + \rho_{xy}^{\mathrm{CA}} + \rho_{xy}^{\mathrm{UA}} = R_0H + R_SM + \rho_{xy}^{\mathrm{UA}}$, is used to disentangle different components. Here, $R_0$ and $R_S$ are normal and conventional anomalous Hall coefficients, respectively while $\rho_{xy}^{\mathrm{UA}}$ is the unconventional component.The positive slope of $\rho_{xy}$ at low $T$ indicates holes-type majority carriers with a density of $n_0 \sim 10^{22}$ cm$^{-3}$, while the intercept yields $\rho_{xy}^{\mathrm{CA}}$(Fig. S3(a) of SM (\cite{SM})). Interestingly, a transition of the carrier-type occurs at 15 K, shifting from holes-like to electrons, before reverting to hole-type at 60 K (Fig. S3(b) in SM\cite{SM}). Figure \ref{F2}(d) depicts the T variation of $\rho_{xy}^\mathrm{CA}$ and anomalous Hall conductivity (AHC), $\sigma_{xy}^\mathrm{CA} \approx \rho_{xy}^\mathrm{CA}/\rho_{xx}^2$, with a substantial value of $\sigma_{xy}^\mathrm{CA}\sim 205$ S cm$^{-1}$ at $T$ = 2 K. It is important to note that (from the Figure \ref{F2}(c)) that an AHE has already been developed in the non-magnetic (NM) phase above $T_C$ (53 K) i.e. typical case of 60, 70 and 100 K $\rho_{xy}$ vs. $H$ curves deviate from the linear behavior above $T_C$. This indicates the onset of short range FM correlations, at a fairly high temperature, which is also evident from our magnetization isotherms above $T_C$  (see Fig. \ref{F1}(e)). In a general, AHE arises from either scattering independent intrinsic or asymmetric extrinsic side-jump ($sj$) or skew-scattering ($sk$) mechanisms or a combination of both\cite{p11}. Consequently, the $\sigma_{xy}^\mathrm{CA}$ can be expresses as $\sigma_{xy}^{CA} = \sigma_{sk} + \sigma_{sj} + \sigma_{int}$, where $\sigma_{int}$ is the intrinsic component. To discern these three contributions, we employed the \textit{TYJ} scaling relation for AHC as, $\sigma_{xy}^{CA} = -a\sigma_{xx0}^{-1}\sigma_{xx}^2 - b $, where $\sigma_{xx0}$ is the residual conductivity and the intercept, $b$, corresponds to combination of $\sigma_{int}$ and $\sigma_{sj}$\cite{p13,p14,p12,p17}. Figure \ref{F2}(e) illustrates the expected linear relation of $\sigma_{xy}^{CA}$ with $\sigma_{xx}^2$ curve, yielding $b \sim 160$ S cm$^{-1}$. \textcolor{black} {Using the power-law relation ($\rho_{xy} \propto \rho_{xx}^\alpha$), along with the parameters $a$ and $b$ from TYJ scaling and the resonance condition, the intrinsic dominance of AHC is confirmed (see Fig. S3 (c) (d)\cite{SM})}\cite{chakraborty2022berry,miyasato2007crossover,onoda2006intrinsic}. Inset of Fig. \ref{F2}(e) illustrates the T-variation of anomalous Hall angle (AHA; $\theta_{AH}$ =$\sigma_{xy}^{CA}/\sigma_{xx}$) and anomalous Hall factor $S_H$ (=$\sigma_{xy}^{CA}/M$). $\theta_{AH}$ attains a value of 2.1$\%$ at 2 K while showing an overall weak T-dependency. While the temperature invariance of $S_H$ at $\sim$0.2 V$^{-1}$ asserts the scattering impervious Karplus-Luttinger origin of AHE \cite{wang2018large,p24,p22,p40}. This quantification of the robust nature of observed conventional AHE, positions $\mathrm{Y}_3\mathrm{Co}_8\mathrm{Sn}_4$ as a compelling material for potential topotronic applications.

Next, we focus on the additional unconventional component of AHE, $\rho_{xy}^{UA}$. Accounting in the intrinsic contribution to AHE, $R_S$ is formulated as $\gamma \rho_{xx}^2$ \cite{p42,p24,p44}. Since any non-collinear arrangements of spin is likely to vanish in the field-polarized state, the intercept of the linearly fitted ($\rho_{xy}/H$) $vs$ ($\rho_{xx}^2M/H$) curves above the critical field ($H_c$) is nothing but $\gamma$. Employing $R_S$, we calculated the ($\rho_{xy}^N + \rho_{xy}^{CA}$) (black curve in Fig. \ref{F2}(c)) and obtained the $\rho_{xy}^{UA}$ using the relation $\rho_{xy}^{UA} = \rho_{xy}- \rho_{xy}^N - R_SM$, at various T, as shown in Fig. \ref{F2}(f). Notably,  $\rho_{xy}^{UA}$ exhibits a qualitative divergence across the transitions in the following way, (i) below 14 K, the field range associated with UAHE decreases with increasing temperature (see inset (i) of Fig. \ref{F2}(f)), while exhibiting an opposite trend above 14 K (inset (ii) of Fig. \ref{F2}(f)). This aligns with the diminished field thresholds for annihilating topologically protected non-trivial structures at elevated thermal fluctuations, while the origin of the UAHE in latter $T$ range (above 14 K) appears to be different. (ii) Alongside, the peak amplitude of $\rho_{xy}^{UA}$ at lower temperatures showcases a decreasing trend from 2 to 14 K, followed by a gradual increase up to 35 K, in contrast to a positive amplitude of $\rho_{xy}^{UA}$  observed below $T_C$ in several topological materials\cite{p34,p35,neubauer2009topological,kumar2020detection}. (iii) the gradual transformation of the positive to negative peak amplitude above $T_C$ is rarely observed. (iv) the negative peak is only observed in NM phase. With the increase of temperature, the positive peak gradually gives way to the
negative peak. For a better visualization, the contour plot of $\rho_{xy}^{UA}(T)$ is overlayed on an $H$-$T$ phase diagram, as illustrated in Fig.\ref{F2}(g). While the nontrivial band topology driven unconventional transverse transport has been observed below and above the magnetic transition temperature in few systems, such as GdPtBi, YbPtBi, and EuCd$_2$As$_2$ \cite{suzuki2016, ybptbi2018, wang2018large}, but to best of our knowledge, this two-stage behavior of the UAHE is uniquely exhibited by the centrosymmetric compound EuCd$_2$As$_2$ \cite{wang2018large}. Interestingly, we observe a similar two-stage behavior for the first time in the non-centrosymmetric compound $\mathrm{Y}_3\mathrm{Co}_8\mathrm{Sn}_4$, underscoring the need for an in-depth investigation into the underlying mechanisms below and above $T_C$. To further elucidate and support our experimental findings, we have performed \textit{ab-initio} density functional theory calculations for both the phases, i.e., below and above $T_C$.

\begin{table*}[t]
	\centering
\caption{Calculated relative energies ($\Delta E$), atom projected Co-moments ($\mu_{Co}$), total magnetic moment ($\mu_{total}$) and average moment per Co atom, for NM, AFM, FiM1 and FiM2 configurations of Y$_3$Co$_8$Sn$_4$. For axial cases, Co-moments are aligned out of plane (indicated by a single $m_z$ component), while for planar cases it lies in-pane (indicated by ($m_x,m_y$) components). There are 6, 6, 2 and 2 Co-atoms on each of the $Co1$, $Co2$, $Co3$ and $Co4$ sublattices (in-equivalent Co-sites) in the unit cell. For AFM case, there are equal number of spin up and down moments denoted by $\pm$ in front of each moment values. While for FiM1 and FiM2, the given moments are for all the (6 or 2) Co-atoms of a given type.}

\begin{ruledtabular}
    	\begin{tabular}{c  c c c c c c c}
			Magnetic & $\Delta{E}$ &  $\mu_{Co1}$  & $\mu_{Co2}$ &  $\mu_{Co3}$ &  $\mu_{Co4}$ & $\mu_{total}$  & Avg. moment/Co  \\

            Configuration & (eV/f.u.) & ($\mu_B$) & ($\mu_B$) &  ($\mu_B$) & ($\mu_B$) & ($\mu_B$/f.u.) & ($\mu_B$)  \\
            \hline	 \\
            NM & 0.667 &  --	& -- & --	& -- & -- & -- \\ \\
                        AFM (Axial) &	0.368	& $\pm$(-0.77)	& $\pm$(0.92)	& $\pm$(-0.05) & $\pm$(0.96) & 0.0 &  0.75 \\ \\
			AFM (Planar) &	0.156	& $\pm$(-0.55,0.55)	& $\pm$(0.65,-0.65)	& $\pm$(-0.03,0.03) & $\pm$(0.68,-0.68) & 0.0 & 0.95 \\ \\
                        
             FiM1 (Axial) &		0.087 & 0.99 & 1.15	& -0.37 & 1.31 & 7.4 & 0.92 \\ \\
            
			FiM1 (Planar) &	0.056	& (0.70,-0.70)	& (0.81,-0.81)	& (-0.26,0.26) & (0.92,-0.92) & 4.9 & 0.61 \\ \\

            FiM2 (Planar) &	0	& (0.70,-0.70)	& (0.80,-0.80)	& (-0.29,0.29) & (-0.92,0.92) & 3.3 & 0.41 \\

             \end{tabular}
             \label{Table1}
\end{ruledtabular}
\end{table*}

To elucidate the electronic structure of $\mathrm{Y}_3\mathrm{Co}_8\mathrm{Sn}_4$ below 53 K, we have simulated various magnetic configurations such as non-magnetic (NM), ferromagnetic (FM), antiferromagnetic (AFM) and ferrimagnetic (FiM1, FiM2), in axial and planar modes (see the schematics in Fig. S4 of SM\cite{SM}), using the experimental structure. For axial cases, the moments are aligned out of plane (i.e. (0,0,$m_z$)) while for planar cases they are aligned in-plane (i.e. ($m_x,m_y,0$)). Interestingly, the system always converges to FiM2 configuration, albeit starting from a FM configuration (with the same spin alignment on all the Co-sites). We have also tried the axial and planar model proposed in Ref. \cite{canepa2004ferromagnetic}, but they also converged back to FiM solution, confirming the robustness of the latter. Table \ref{Table1} enlists the simulated relative energies, total and Co-projected magnetic moments, and average moment per Co atom for all the distinct converged magnetic orderings. Clearly, the FiM2 ordering (with Co moments oriented on easy \textit{ab}-plane) is found to be energetically the most stable configuration. \textit{ab}-plane has been reported to be the easy plane in all the experimental reports so far \cite{canepa2004ferromagnetic, takagi2018multiple}. The robustness of in-plane FiM ordering is also evident from the experimental non-saturating M-H curve below 53 K (shown in Fig. 1(e)).  In FiM2 configuration, $Co3$ (2b site) and $Co4$ (2a site) atoms are antiferromagnetically coupled to the other Co-atoms, which gives rise to a net magnetization of around 3.3 $\mu_{B}$/f.u. The average magnetic moment per Co-atom is 0.41 $\mu_{B}$/f.u., agreeing reasonably well with our experimental result.

\begin{figure}[t]
\begin{center}
\includegraphics[width=.485\textwidth]{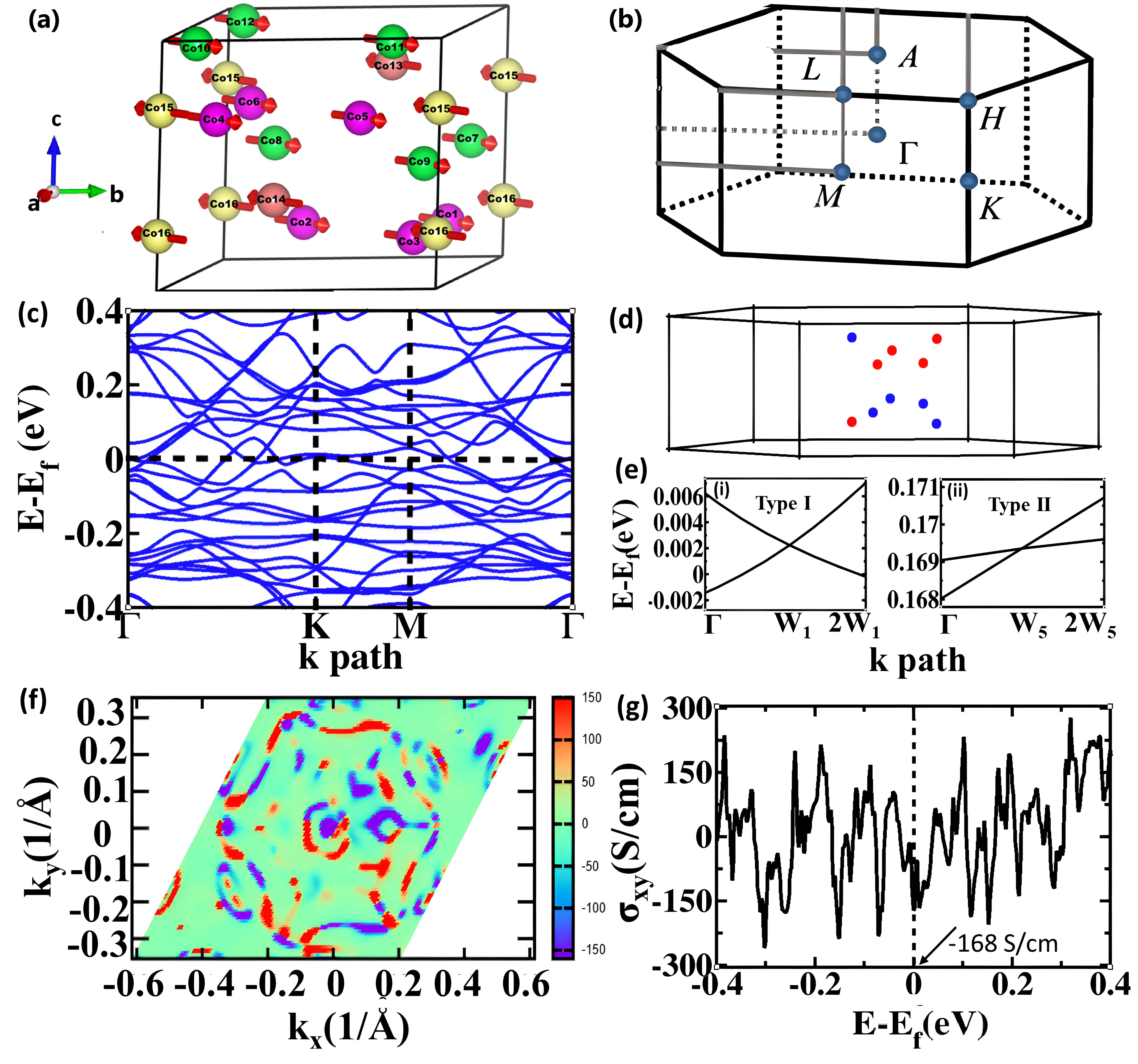}
\end{center}
\caption[]
{\label{Fig.3}
For $\mathrm{Y}_3\mathrm{Co}_8\mathrm{Sn}_4$, (a) Unit cell schematic along with lowest energy planar FiM2 magnetic ordering. Only Co atoms at different sublattices (i.e. $Co1, Co2, Co3, Co4$) along with their moment directions are shown, while Y and Sn atoms are omitted for better visualization. (b) Hexagonal bulk Brillouin zone (BZ) (c) Electronic bulk band structure including SOC (d) Distribution of 5 pairs of WPs in the bulk BZ.  WPs with +1(-1) chirality are shown by red (blue) dots. (e) Bulk band dispersion in the vicinity of a representative (i) type-I and (ii) type-II WPs. (f) Berry curvature in the $k_x$-$k_y$ plane. (g) Energy dependence of anomalous Hall conductivity ($\sigma_{xy}$). }
\label{fig3}
\end{figure}

Figure \ref{Fig.3}(a) shows the schematic of unit cell with lowest energy planar FiM2 magnetic ordering (only Co-atoms are displayed for clarity).
Figure \ref{Fig.3}(c) shows the electronic bulk band structure of $\mathrm{Y}_3\mathrm{Co}_8\mathrm{Sn}_4$ in this ordering with spin-orbit coupling (SOC), using the theoretically optimized lattice parameters (mentioned in SM\cite{SM}). The metallic nature is evident with several linear crossings around the Fermi level (E$_F$). Inclusion of SOC does not change the band topology much, but induce small splitting  along these linear crossings\cite{wang2018large},\cite{zrsis2016},\cite{capd2018} (see band structure without SOC in Fig. S5 of SM\cite{SM}). Co \textit{d}-orbitals show major contributions to the density of states (DoS) at/near the E$_F$.
To investigate the topological features of the band structure, we performed an extensive search of topological nodal points in the entire bulk Brillouin zone (BZ) (schematic shown in Fig. \ref{Fig.3}(b)). This yields a total of 5 pairs of WPs with +/- chirality, as shown in Fig. \ref{Fig.3}(d). The reciprocal space coordinates, types, energy positions (with respect to E$_F$) and the chirality of these WPs are shown in Table \ref{Table2}. Out of these 5 pairs of WPs, 3 are of type-I, while the other 2 are of type-II nature. One can notice from the table that most of these WPs lie very close to the E$_F$, leading to the possibility of large Berry curvature, and hence the intrinsic anomalous Hall conductivity, as reported for many similar systems such as Kagome ferromagnet Fe$_3$Sn$_2$ \cite{fe3sn22011}, noncollinear antiferromagnet Mn$_3$Sn \cite{mn3sn2015}, magnetic Weyl semimetal Co$_3$Sn$_3$S$_2$ \cite{wang2018large} etc.  A zoomed-in view of the bulk band dispersion near the crossing points  for a representative type-I and type-II WPs are shown in Fig. \ref{Fig.3}(e). Intrinsic AHE are intimately connected to the Berry curvature ($\Omega_n(k)$) of occupied electronic states, arising out of the WPs. Figure \ref{Fig.3}(f) shows the Berry curvature on the $k_x$-$k_y$ plane. The color schema facilitates to visualize the hot spots in the $\Omega_n(k)$ values.  AHC ($\sigma_{xy}$) can be calculated from ($\Omega_n(k)$) using the following relation, $\sigma_{xy}= - e^2/\hbar \int_{BZ} d^3k/(2\pi)^3 \sum_{n} f_n(k) \Omega_{n}^{z}(k)$, where $f_n(k)$ is Fermi–Dirac distribution function, $n$ is the occupied bands index, and $\Omega_{n}^{z}(k)$  is the z-component of Berry curvature \cite{wang2018large}. 
Figure \ref{Fig.3}(g) shows the energy dependence of AHC for FiM2 configuration of $\mathrm{Y}_3\mathrm{Co}_8\mathrm{Sn}_4$. The absolute value of AHC is found to be around 168 S/cm at E$_F$, which is in fair agreement with our experimental value $\sim$160 S/cm at 30 K (see Fig. S3(d)\cite{SM}) for $\mathrm{Y}_3\mathrm{Co}_8\mathrm{Sn}_4$ below 53 K. This clearly shows that there is a predominant intrinsic contribution to AHE in the present compound below 53 K.

\begin{table}[t]
	\centering
\caption{For FiM2 ordering of Y$_3$Co$_8$Sn$_4$, reciprocal space coordinates (in \AA$^{-1}$), type, energy position relative to E$_F$ (in eV) and chirality of the five pairs of Weyl points (WPs).}
\label{Table2}
\begin{ruledtabular}
    	\begin{tabular}{c  c c c c}
			WPs &  Coordinates & Type  & Energy  & Chirality  \\
             &  of WP  &   & position  &   \\
            \hline	
            $W_{1}^{-}$ & (-0.229,-0.299,0.200) & I & 0.003 & -1 \\
           $W_{1}^{+}$ & (0.226,-0.298,-0.200 & I & 0.002 & +1 \\
           $W_{2}^{-}$ & (-0.066,0.059,-0.308) & I & 0.140 & -1 \\
           $W_{2}^{+}$ & (0.069,0.057,0.303) & I & 0.130 & +1 \\
           $W_{3}^{-}$ &  (0.352,0.004,-0.421) & I & 0.075 & -1 \\
           $W_{3}^{+}$ &  (0.355,-0.002,0.421) & I & 0.075 & +1 \\
           $W_{4}^{-}$ & (-0.177,0.310,0.420) & II & 0.073 & -1 \\
           $W_{4}^{+}$ & (-0.176,0.320,-0.419) & II & 0.074 & +1 \\
           $W_{5}^{-}$ & (0.040,-0.044,-0.167) & II & 0.169 & -1 \\
           $W_{5}^{+}$ & (-0.040,0.043,0.166) & II & 0.169 & +1 \\
			 
             \end{tabular}
\end{ruledtabular}
\end{table}

 To investigate the unconventional transport properties and their mechanism above 53 K, we have also simulated the electronic  properties of Y$_3$Co$_8$Sn$_4$ in the nonmagnetic phase. The electronic bulk band structure of the present compound without and with SOC in nonmagnetic configuration, is shown in Fig. S6 and S7(a) in SM\cite{SM} respectively. It shows a metallic nature, with few linear crossings around the E$_F$, which are to be checked further for their trivial/nontrivial topological nature. It is interesting to note here that the inclusion of SOC leads to significant changes in the band topology, including large energy gap opening around some of the linear crossings, which resembles the outcome for other similar topological materials\cite{socgapnacabi,linearresponse2019}.

As reported earlier, the finite UAHE above $T_C$ can only be attributed to momentum-space lead Berry curvature arising out of the WPs \cite{wang2018large}. Therefore, we further checked the nature of these linear crossings in the nonmagnetic phase of Y$_3$Co$_8$Sn$_4$. For that, a search of topological nodal points in the entire bulk BZ is carried out, which revealed a total of 4 pairs of WPs with opposite chirality, as shown by red and blue dots in the bulk BZ in Fig. S7(b)\cite{SM}. Of these 4 pairs, 1 pair is of type-I, whereas the other 3  are of type-II nature. A zoomed-in view of the band structure in the vicinity of a representative type I and type II Weyl crossings are shown in Fig. S7(c,d)\cite{SM}. Total topological charge associated with these 8 WPs sum up to zero, which is found to be in accordance with the Nielsen-Ninomiya theorem\cite{ninomiya1981}. Reciprocal space coordinates, type, energy position (relative to E$_F$) and chirality of these 4 pairs of WPs are shown in Table ST2 of SM\cite{SM}. One can see from the table that most of these WPs lie close to E$_F$, which leads to a large Berry curvature as presented in $k_x$-$k_y$ plane in Fig. S7(e). 

To investigate the intrinsic contribution to AHE  in the case of non-magnetic phase (above 53 K), we have calculated the AHC in a wide energy range of E$_F \pm$0.4 eV, as shown in Fig. S7(f)\cite{SM}. The absolute value of AHC is found to be around 154 S/cm at E$_F$, which agrees fairly well with the measured value for typical case of 60 K (i.e. above $T_C$) (136 S cm$^{-1}$ (see Fig. S3(d)\cite{SM})). This suggests that the main contribution to the observed AHE above 53 K in the nonmagnetic phase is also arising from the intrinsic mechanism, which is intimately related to the occurrence of WPS at/near $E_F$. Such observation in the nonmagnetic phase can be understood using a simple idea as follows. At higher temperatures, short range FM domains may exist in the system, where each domain has the magnetic moments in the same direction. But these FM domains are oriented randomly, and as a whole it gives rise to the non-magnetic phase. In each FM domain, WPs may occur due to the breaking of time-reversal symmetry. However, the pair polarity of each WPs is compensated by that of the other WPs pair present in another domain, having opposite magnetization.
Thus the negative and positive contribution to transverse resistivity would be canceled out, explaining the lack of  spontaneous component of UAHE in NM phase.

{\it \bf Conclusion:} In conclusion, this study reveals the interplay between magnetism and topology in a non-centrosymmetric compound $\mathrm{Y}_3\mathrm{Co}_8\mathrm{Sn}_4$, highlighting its unique magnetic and transport properties. Below 14 K, complex spin textures are evident, accompanied by a sharp rise in resistivity and large negative magnetoresistance. The dominant intrinsic anomalous Hall conductivity and an unconventional anomalous Hall signature, with a two-stage behavior—positive below $T_C$ and negative above $T_C$—stem from a combination of intrinsic band topology, and magnetism, such as an unique FiM state below $T_C$, and short-range FM fluctuations present above $T_C$. $Ab$-$initio$ calculations reveal the existence of topological Weyl points in both in-plane ferrimagnetic (below $T_C$) and non-magnetic (above $T_C$) states, underscoring the role of intrinsic Berry curvature in UAHE. A reasonably large value of AHC has also been observed in both the phases (below and above $T_C$) in the present compound, which is rarely observed in a single same system. These findings establish $\mathrm{Y}_3\mathrm{Co}_8\mathrm{Sn}_4$ as a potential platform to further explore topological magnetism with promising applications in spintronics and quantum transport technologies.

\textbf{Acknowledgement:} A.A. and A.B. gratefully acknowledge SINP, India, and the Department of Atomic Energy (DAE), India, for their fellowships. Work at the Ames National Laboratory was supported by the Division of Materials Science and Engineering, Office of Basic Energy Sciences, Office of Science, U.S. Department of Energy (DOE). Ames National Laboratory is operated for the U.S. DOE by Iowa State University of Science and Technology under Contract No. DE-AC02-07CH11358.

\bibliography{ref}

\begin{thebibliography}{68}%
\makeatletter
\providecommand \@ifxundefined [1]{%
 \@ifx{#1\undefined}
}%
\providecommand \@ifnum [1]{%
 \ifnum #1\expandafter \@firstoftwo
 \else \expandafter \@secondoftwo
 \fi
}%
\providecommand \@ifx [1]{%
 \ifx #1\expandafter \@firstoftwo
 \else \expandafter \@secondoftwo
 \fi
}%
\providecommand \natexlab [1]{#1}%
\providecommand \enquote  [1]{``#1''}%
\providecommand \bibnamefont  [1]{#1}%
\providecommand \bibfnamefont [1]{#1}%
\providecommand \citenamefont [1]{#1}%
\providecommand \href@noop [0]{\@secondoftwo}%
\providecommand \href [0]{\begingroup \@sanitize@url \@href}%
\providecommand \@href[1]{\@@startlink{#1}\@@href}%
\providecommand \@@href[1]{\endgroup#1\@@endlink}%
\providecommand \@sanitize@url [0]{\catcode `\\12\catcode `\$12\catcode
  `\&12\catcode `\#12\catcode `\^12\catcode `\_12\catcode `\%12\relax}%
\providecommand \@@startlink[1]{}%
\providecommand \@@endlink[0]{}%
\providecommand \url  [0]{\begingroup\@sanitize@url \@url }%
\providecommand \@url [1]{\endgroup\@href {#1}{\urlprefix }}%
\providecommand \urlprefix  [0]{URL }%
\providecommand \Eprint [0]{\href }%
\@ifxundefined \urlstyle {%
  \providecommand \doi  [0]{\begingroup \@sanitize@url \@doi}%
  \providecommand \@doi [1]{\endgroup \@@startlink {\doibase
  #1}doi:\discretionary {}{}{}#1\@@endlink }%
}{%
  \providecommand \doi  [0]{doi:\discretionary{}{}{}\begingroup
  \urlstyle{rm}\Url }%
}%
\providecommand \doibase [0]{http://dx.doi.org/}%
\providecommand \Doi [0]{\begingroup \@sanitize@url \@Doi }%
\providecommand \@Doi  [1]{\endgroup\@@startlink{\doibase#1}\@@Doi}%
\providecommand \@@Doi [1]{#1\@@endlink}%
\providecommand \selectlanguage [0]{\@gobble}%
\providecommand \bibinfo  [0]{\@secondoftwo}%
\providecommand \bibfield  [0]{\@secondoftwo}%
\providecommand \translation [1]{[#1]}%
\providecommand \BibitemOpen [0]{}%
\providecommand \bibitemStop [0]{}%
\providecommand \bibitemNoStop [0]{.\EOS\space}%
\providecommand \EOS [0]{\spacefactor3000\relax}%
\providecommand \BibitemShut  [1]{\csname bibitem#1\endcsname}%
\bibitem [{\citenamefont {Liu}\ \emph {et~al.}(2019)\citenamefont {Liu},
  \citenamefont {Liang}, \citenamefont {Liu}, \citenamefont {Xu}, \citenamefont
  {Li}, \citenamefont {Chen}, \citenamefont {Pei}, \citenamefont {Shi},
  \citenamefont {Mo}, \citenamefont {Dudin}, \citenamefont {Kim}, \citenamefont
  {Cacho}, \citenamefont {Li}, \citenamefont {Sun}, \citenamefont {Yang},
  \citenamefont {Liu}, \citenamefont {Parkin}, \citenamefont {Felser},\ and\
  \citenamefont {Chen}}]{science.aav2873}%
  \BibitemOpen
  \bibfield  {author} {\bibinfo {author} {\bibfnamefont {D.~F.}\ \bibnamefont
  {Liu}}, \bibinfo {author} {\bibfnamefont {A.~J.}\ \bibnamefont {Liang}},
  \bibinfo {author} {\bibfnamefont {E.~K.}\ \bibnamefont {Liu}}, \bibinfo
  {author} {\bibfnamefont {Q.~N.}\ \bibnamefont {Xu}}, \bibinfo {author}
  {\bibfnamefont {Y.~W.}\ \bibnamefont {Li}}, \bibinfo {author} {\bibfnamefont
  {C.}~\bibnamefont {Chen}}, \bibinfo {author} {\bibfnamefont {D.}~\bibnamefont
  {Pei}}, \bibinfo {author} {\bibfnamefont {W.~J.}\ \bibnamefont {Shi}},
  \bibinfo {author} {\bibfnamefont {S.~K.}\ \bibnamefont {Mo}}, \bibinfo
  {author} {\bibfnamefont {P.}~\bibnamefont {Dudin}}, \bibinfo {author}
  {\bibfnamefont {T.}~\bibnamefont {Kim}}, \bibinfo {author} {\bibfnamefont
  {C.}~\bibnamefont {Cacho}}, \bibinfo {author} {\bibfnamefont
  {G.}~\bibnamefont {Li}}, \bibinfo {author} {\bibfnamefont {Y.}~\bibnamefont
  {Sun}}, \bibinfo {author} {\bibfnamefont {L.~X.}\ \bibnamefont {Yang}},
  \bibinfo {author} {\bibfnamefont {Z.~K.}\ \bibnamefont {Liu}}, \bibinfo
  {author} {\bibfnamefont {S.~S.~P.}\ \bibnamefont {Parkin}}, \bibinfo {author}
  {\bibfnamefont {C.}~\bibnamefont {Felser}}, \ and\ \bibinfo {author}
  {\bibfnamefont {Y.~L.}\ \bibnamefont {Chen}},\ }\href@noop {} {\bibfield
  {journal} {\bibinfo  {journal} {Sci.},\ }\textbf {\bibinfo {volume} {365}},\
  \bibinfo {pages} {1282} (\bibinfo {year} {2019})}\BibitemShut {NoStop}%
\bibitem [{\citenamefont {Keimer}\ and\ \citenamefont
  {Moore}(2017)}]{keimer2017physics}%
  \BibitemOpen
  \bibfield  {author} {\bibinfo {author} {\bibfnamefont {B.}~\bibnamefont
  {Keimer}}\ and\ \bibinfo {author} {\bibfnamefont {J.}~\bibnamefont {Moore}},\
  }\href@noop {} {\bibfield  {journal} {\bibinfo  {journal} {Nat. Phys.},\
  }\textbf {\bibinfo {volume} {13}},\ \bibinfo {pages} {1045} (\bibinfo {year}
  {2017})}\BibitemShut {NoStop}%
\bibitem [{\citenamefont {Belopolski}\ \emph {et~al.}(2019)\citenamefont
  {Belopolski}, \citenamefont {Manna}, \citenamefont {Sanchez}, \citenamefont
  {Chang}, \citenamefont {Ernst}, \citenamefont {Yin}, \citenamefont {Zhang},
  \citenamefont {Cochran}, \citenamefont {Shumiya}, \citenamefont {Zheng} \emph
  {et~al.}}]{belopolski2019discovery}%
  \BibitemOpen
  \bibfield  {author} {\bibinfo {author} {\bibfnamefont {I.}~\bibnamefont
  {Belopolski}}, \bibinfo {author} {\bibfnamefont {K.}~\bibnamefont {Manna}},
  \bibinfo {author} {\bibfnamefont {D.~S.}\ \bibnamefont {Sanchez}}, \bibinfo
  {author} {\bibfnamefont {G.}~\bibnamefont {Chang}}, \bibinfo {author}
  {\bibfnamefont {B.}~\bibnamefont {Ernst}}, \bibinfo {author} {\bibfnamefont
  {J.}~\bibnamefont {Yin}}, \bibinfo {author} {\bibfnamefont {S.~S.}\
  \bibnamefont {Zhang}}, \bibinfo {author} {\bibfnamefont {T.}~\bibnamefont
  {Cochran}}, \bibinfo {author} {\bibfnamefont {N.}~\bibnamefont {Shumiya}},
  \bibinfo {author} {\bibfnamefont {H.}~\bibnamefont {Zheng}},  \emph
  {et~al.},\ }\href@noop {} {\bibfield  {journal} {\bibinfo  {journal} {Sci.},\
  }\textbf {\bibinfo {volume} {365}},\ \bibinfo {pages} {1278} (\bibinfo {year}
  {2019})}\BibitemShut {NoStop}%
\bibitem [{\citenamefont {Wang}\ \emph {et~al.}(2018)\citenamefont {Wang},
  \citenamefont {Xu}, \citenamefont {Lou}, \citenamefont {Liu}, \citenamefont
  {Li}, \citenamefont {Huang}, \citenamefont {Shen}, \citenamefont {Weng},
  \citenamefont {Wang},\ and\ \citenamefont {Lei}}]{wang2018large}%
  \BibitemOpen
  \bibfield  {author} {\bibinfo {author} {\bibfnamefont {Q.}~\bibnamefont
  {Wang}}, \bibinfo {author} {\bibfnamefont {Y.}~\bibnamefont {Xu}}, \bibinfo
  {author} {\bibfnamefont {R.}~\bibnamefont {Lou}}, \bibinfo {author}
  {\bibfnamefont {Z.}~\bibnamefont {Liu}}, \bibinfo {author} {\bibfnamefont
  {M.}~\bibnamefont {Li}}, \bibinfo {author} {\bibfnamefont {Y.}~\bibnamefont
  {Huang}}, \bibinfo {author} {\bibfnamefont {D.}~\bibnamefont {Shen}},
  \bibinfo {author} {\bibfnamefont {H.}~\bibnamefont {Weng}}, \bibinfo {author}
  {\bibfnamefont {S.}~\bibnamefont {Wang}}, \ and\ \bibinfo {author}
  {\bibfnamefont {H.}~\bibnamefont {Lei}},\ }\href@noop {} {\bibfield
  {journal} {\bibinfo  {journal} {Nat. Commun.},\ }\textbf {\bibinfo {volume}
  {9}},\ \bibinfo {pages} {1} (\bibinfo {year} {2018})}\BibitemShut {NoStop}%
\bibitem [{\citenamefont {Hasan}\ and\ \citenamefont
  {Kane}(2010)}]{hasan2010colloquium}%
  \BibitemOpen
  \bibfield  {author} {\bibinfo {author} {\bibfnamefont {M.~Z.}\ \bibnamefont
  {Hasan}}\ and\ \bibinfo {author} {\bibfnamefont {C.~L.}\ \bibnamefont
  {Kane}},\ }\href@noop {} {\bibfield  {journal} {\bibinfo  {journal} {Rev.
  Mod. Phys.},\ }\textbf {\bibinfo {volume} {82}},\ \bibinfo {pages} {3045}
  (\bibinfo {year} {2010})}\BibitemShut {NoStop}%
\bibitem [{\citenamefont {Neubauer}\ \emph {et~al.}(2009)\citenamefont
  {Neubauer}, \citenamefont {Pfleiderer}, \citenamefont {Binz}, \citenamefont
  {Rosch}, \citenamefont {Ritz}, \citenamefont {Niklowitz},\ and\ \citenamefont
  {B{\"o}ni}}]{neubauer2009topological}%
  \BibitemOpen
  \bibfield  {author} {\bibinfo {author} {\bibfnamefont {A.}~\bibnamefont
  {Neubauer}}, \bibinfo {author} {\bibfnamefont {C.}~\bibnamefont
  {Pfleiderer}}, \bibinfo {author} {\bibfnamefont {B.}~\bibnamefont {Binz}},
  \bibinfo {author} {\bibfnamefont {A.}~\bibnamefont {Rosch}}, \bibinfo
  {author} {\bibfnamefont {R.}~\bibnamefont {Ritz}}, \bibinfo {author}
  {\bibfnamefont {P.}~\bibnamefont {Niklowitz}}, \ and\ \bibinfo {author}
  {\bibfnamefont {P.}~\bibnamefont {B{\"o}ni}},\ }\href@noop {} {\bibfield
  {journal} {\bibinfo  {journal} {Phys. Rev. Lett.},\ }\textbf {\bibinfo
  {volume} {102}},\ \bibinfo {pages} {186602} (\bibinfo {year}
  {2009})}\BibitemShut {NoStop}%
\bibitem [{\citenamefont {Fert}\ \emph {et~al.}(2017)\citenamefont {Fert},
  \citenamefont {Reyren},\ and\ \citenamefont {Cros}}]{fert2017magnetic}%
  \BibitemOpen
  \bibfield  {author} {\bibinfo {author} {\bibfnamefont {A.}~\bibnamefont
  {Fert}}, \bibinfo {author} {\bibfnamefont {N.}~\bibnamefont {Reyren}}, \ and\
  \bibinfo {author} {\bibfnamefont {V.}~\bibnamefont {Cros}},\ }\href@noop {}
  {\bibfield  {journal} {\bibinfo  {journal} {Nat. Rev. Mater.},\ }\textbf
  {\bibinfo {volume} {2}},\ \bibinfo {pages} {1} (\bibinfo {year}
  {2017})}\BibitemShut {NoStop}%
\bibitem [{\citenamefont {Heinze}\ \emph {et~al.}(2011)\citenamefont {Heinze},
  \citenamefont {Von~Bergmann}, \citenamefont {Menzel}, \citenamefont {Brede},
  \citenamefont {Kubetzka}, \citenamefont {Wiesendanger}, \citenamefont
  {Bihlmayer},\ and\ \citenamefont {Bl{\"u}gel}}]{heinze2011spontaneous}%
  \BibitemOpen
  \bibfield  {author} {\bibinfo {author} {\bibfnamefont {S.}~\bibnamefont
  {Heinze}}, \bibinfo {author} {\bibfnamefont {K.}~\bibnamefont
  {Von~Bergmann}}, \bibinfo {author} {\bibfnamefont {M.}~\bibnamefont
  {Menzel}}, \bibinfo {author} {\bibfnamefont {J.}~\bibnamefont {Brede}},
  \bibinfo {author} {\bibfnamefont {A.}~\bibnamefont {Kubetzka}}, \bibinfo
  {author} {\bibfnamefont {R.}~\bibnamefont {Wiesendanger}}, \bibinfo {author}
  {\bibfnamefont {G.}~\bibnamefont {Bihlmayer}}, \ and\ \bibinfo {author}
  {\bibfnamefont {S.}~\bibnamefont {Bl{\"u}gel}},\ }\href@noop {} {\bibfield
  {journal} {\bibinfo  {journal} {Nat. Phys.},\ }\textbf {\bibinfo {volume}
  {7}},\ \bibinfo {pages} {713} (\bibinfo {year} {2011})}\BibitemShut {NoStop}%
\bibitem [{\citenamefont {Shekhar}\ \emph {et~al.}(2018)\citenamefont
  {Shekhar}, \citenamefont {Kumar}, \citenamefont {Grinenko}, \citenamefont
  {Singh}, \citenamefont {Sarkar}, \citenamefont {Luetkens}, \citenamefont
  {Wu}, \citenamefont {Zhang}, \citenamefont {Komarek}, \citenamefont {Kampert}
  \emph {et~al.}}]{shekhar2018anomalous}%
  \BibitemOpen
  \bibfield  {author} {\bibinfo {author} {\bibfnamefont {C.}~\bibnamefont
  {Shekhar}}, \bibinfo {author} {\bibfnamefont {N.}~\bibnamefont {Kumar}},
  \bibinfo {author} {\bibfnamefont {V.}~\bibnamefont {Grinenko}}, \bibinfo
  {author} {\bibfnamefont {S.}~\bibnamefont {Singh}}, \bibinfo {author}
  {\bibfnamefont {R.}~\bibnamefont {Sarkar}}, \bibinfo {author} {\bibfnamefont
  {H.}~\bibnamefont {Luetkens}}, \bibinfo {author} {\bibfnamefont {S.-C.}\
  \bibnamefont {Wu}}, \bibinfo {author} {\bibfnamefont {Y.}~\bibnamefont
  {Zhang}}, \bibinfo {author} {\bibfnamefont {A.~C.}\ \bibnamefont {Komarek}},
  \bibinfo {author} {\bibfnamefont {E.}~\bibnamefont {Kampert}},  \emph
  {et~al.},\ }\href@noop {} {\bibfield  {journal} {\bibinfo  {journal}
  {PNAS.},\ }\textbf {\bibinfo {volume} {115}},\ \bibinfo {pages} {9140}
  (\bibinfo {year} {2018})}\BibitemShut {NoStop}%
\bibitem [{\citenamefont {Burkov}(2014)}]{burkov2014anomalous}%
  \BibitemOpen
  \bibfield  {author} {\bibinfo {author} {\bibfnamefont {A.}~\bibnamefont
  {Burkov}},\ }\href@noop {} {\bibfield  {journal} {\bibinfo  {journal} {Phys.
  Rev. Lett.},\ }\textbf {\bibinfo {volume} {113}},\ \bibinfo {pages} {187202}
  (\bibinfo {year} {2014})}\BibitemShut {NoStop}%
\bibitem [{\citenamefont {Cheng}\ \emph {et~al.}(2024)\citenamefont {Cheng},
  \citenamefont {Yan}, \citenamefont {Shi}, \citenamefont {Lou}, \citenamefont
  {Fedorov}, \citenamefont {Behnami}, \citenamefont {Yuan}, \citenamefont
  {Yang}, \citenamefont {Wang}, \citenamefont {Cheng} \emph
  {et~al.}}]{cheng2024tunable}%
  \BibitemOpen
  \bibfield  {author} {\bibinfo {author} {\bibfnamefont {E.}~\bibnamefont
  {Cheng}}, \bibinfo {author} {\bibfnamefont {L.}~\bibnamefont {Yan}}, \bibinfo
  {author} {\bibfnamefont {X.}~\bibnamefont {Shi}}, \bibinfo {author}
  {\bibfnamefont {R.}~\bibnamefont {Lou}}, \bibinfo {author} {\bibfnamefont
  {A.}~\bibnamefont {Fedorov}}, \bibinfo {author} {\bibfnamefont
  {M.}~\bibnamefont {Behnami}}, \bibinfo {author} {\bibfnamefont
  {J.}~\bibnamefont {Yuan}}, \bibinfo {author} {\bibfnamefont {P.}~\bibnamefont
  {Yang}}, \bibinfo {author} {\bibfnamefont {B.}~\bibnamefont {Wang}}, \bibinfo
  {author} {\bibfnamefont {J.-G.}\ \bibnamefont {Cheng}},  \emph {et~al.},\
  }\href@noop {} {\bibfield  {journal} {\bibinfo  {journal} {Nat. Commun.},\
  }\textbf {\bibinfo {volume} {15}},\ \bibinfo {pages} {1467} (\bibinfo {year}
  {2024})}\BibitemShut {NoStop}%
\bibitem [{\citenamefont {Xu}\ \emph {et~al.}(2021)\citenamefont {Xu},
  \citenamefont {Das}, \citenamefont {Ma}, \citenamefont {Yi}, \citenamefont
  {Nie}, \citenamefont {Shi}, \citenamefont {Tiwari}, \citenamefont {Tsirkin},
  \citenamefont {Neupert}, \citenamefont {Medarde} \emph
  {et~al.}}]{xu2021unconventional}%
  \BibitemOpen
  \bibfield  {author} {\bibinfo {author} {\bibfnamefont {Y.}~\bibnamefont
  {Xu}}, \bibinfo {author} {\bibfnamefont {L.}~\bibnamefont {Das}}, \bibinfo
  {author} {\bibfnamefont {J.}~\bibnamefont {Ma}}, \bibinfo {author}
  {\bibfnamefont {C.}~\bibnamefont {Yi}}, \bibinfo {author} {\bibfnamefont
  {S.}~\bibnamefont {Nie}}, \bibinfo {author} {\bibfnamefont {Y.}~\bibnamefont
  {Shi}}, \bibinfo {author} {\bibfnamefont {A.}~\bibnamefont {Tiwari}},
  \bibinfo {author} {\bibfnamefont {S.~S.}\ \bibnamefont {Tsirkin}}, \bibinfo
  {author} {\bibfnamefont {T.}~\bibnamefont {Neupert}}, \bibinfo {author}
  {\bibfnamefont {M.}~\bibnamefont {Medarde}},  \emph {et~al.},\ }\href@noop {}
  {\bibfield  {journal} {\bibinfo  {journal} {Phys. Rev. Lett.},\ }\textbf
  {\bibinfo {volume} {126}},\ \bibinfo {pages} {076602} (\bibinfo {year}
  {2021})}\BibitemShut {NoStop}%
\bibitem [{\citenamefont {Du}\ \emph {et~al.}(2022)\citenamefont {Du},
  \citenamefont {Hu}, \citenamefont {Han}, \citenamefont {Camino},
  \citenamefont {Zhu},\ and\ \citenamefont
  {Petrovic}}]{PhysRevLett.129.236601}%
  \BibitemOpen
  \bibfield  {author} {\bibinfo {author} {\bibfnamefont {Q.}~\bibnamefont
  {Du}}, \bibinfo {author} {\bibfnamefont {Z.}~\bibnamefont {Hu}}, \bibinfo
  {author} {\bibfnamefont {M.-G.}\ \bibnamefont {Han}}, \bibinfo {author}
  {\bibfnamefont {F.}~\bibnamefont {Camino}}, \bibinfo {author} {\bibfnamefont
  {Y.}~\bibnamefont {Zhu}}, \ and\ \bibinfo {author} {\bibfnamefont
  {C.}~\bibnamefont {Petrovic}},\ }\href@noop {} {\bibfield  {journal}
  {\bibinfo  {journal} {Phys. Rev. Lett.},\ }\textbf {\bibinfo {volume}
  {129}},\ \bibinfo {pages} {236601} (\bibinfo {year} {2022})}\BibitemShut
  {NoStop}%
\bibitem [{\citenamefont {Puphal}\ \emph {et~al.}(2020)\citenamefont {Puphal},
  \citenamefont {Pomjakushin}, \citenamefont {Kanazawa}, \citenamefont
  {Ukleev}, \citenamefont {Gawryluk}, \citenamefont {Ma}, \citenamefont
  {Naamneh}, \citenamefont {Plumb}, \citenamefont {Keller}, \citenamefont
  {Cubitt}, \citenamefont {Pomjakushina},\ and\ \citenamefont
  {White}}]{PhysRevLett.124.017202}%
  \BibitemOpen
  \bibfield  {author} {\bibinfo {author} {\bibfnamefont {P.}~\bibnamefont
  {Puphal}}, \bibinfo {author} {\bibfnamefont {V.}~\bibnamefont {Pomjakushin}},
  \bibinfo {author} {\bibfnamefont {N.}~\bibnamefont {Kanazawa}}, \bibinfo
  {author} {\bibfnamefont {V.}~\bibnamefont {Ukleev}}, \bibinfo {author}
  {\bibfnamefont {D.~J.}\ \bibnamefont {Gawryluk}}, \bibinfo {author}
  {\bibfnamefont {J.}~\bibnamefont {Ma}}, \bibinfo {author} {\bibfnamefont
  {M.}~\bibnamefont {Naamneh}}, \bibinfo {author} {\bibfnamefont {N.~C.}\
  \bibnamefont {Plumb}}, \bibinfo {author} {\bibfnamefont {L.}~\bibnamefont
  {Keller}}, \bibinfo {author} {\bibfnamefont {R.}~\bibnamefont {Cubitt}},
  \bibinfo {author} {\bibfnamefont {E.}~\bibnamefont {Pomjakushina}}, \ and\
  \bibinfo {author} {\bibfnamefont {J.~S.}\ \bibnamefont {White}},\ }\href@noop
  {} {\bibfield  {journal} {\bibinfo  {journal} {Phys. Rev. Lett.},\ }\textbf
  {\bibinfo {volume} {124}},\ \bibinfo {pages} {017202} (\bibinfo {year}
  {2020})}\BibitemShut {NoStop}%
\bibitem [{\citenamefont {Guo}\ \emph {et~al.}(2018)\citenamefont {Guo},
  \citenamefont {Wu}, \citenamefont {Wu}, \citenamefont {Smidman},
  \citenamefont {Cao}, \citenamefont {Bostwick}, \citenamefont {Jozwiak},
  \citenamefont {Rotenberg}, \citenamefont {Liu}, \citenamefont {Steglich}
  \emph {et~al.}}]{ybptbi2018}%
  \BibitemOpen
  \bibfield  {author} {\bibinfo {author} {\bibfnamefont {C.}~\bibnamefont
  {Guo}}, \bibinfo {author} {\bibfnamefont {F.}~\bibnamefont {Wu}}, \bibinfo
  {author} {\bibfnamefont {Z.}~\bibnamefont {Wu}}, \bibinfo {author}
  {\bibfnamefont {M.}~\bibnamefont {Smidman}}, \bibinfo {author} {\bibfnamefont
  {C.}~\bibnamefont {Cao}}, \bibinfo {author} {\bibfnamefont {A.}~\bibnamefont
  {Bostwick}}, \bibinfo {author} {\bibfnamefont {C.}~\bibnamefont {Jozwiak}},
  \bibinfo {author} {\bibfnamefont {E.}~\bibnamefont {Rotenberg}}, \bibinfo
  {author} {\bibfnamefont {Y.}~\bibnamefont {Liu}}, \bibinfo {author}
  {\bibfnamefont {F.}~\bibnamefont {Steglich}},  \emph {et~al.},\ }\href@noop
  {} {\bibfield  {journal} {\bibinfo  {journal} {Nat. commun.},\ }\textbf
  {\bibinfo {volume} {9}},\ \bibinfo {pages} {4622} (\bibinfo {year}
  {2018})}\BibitemShut {NoStop}%
\bibitem [{\citenamefont {Soluyanov}\ \emph {et~al.}(2015)\citenamefont
  {Soluyanov}, \citenamefont {Gresch}, \citenamefont {Wang}, \citenamefont
  {Wu}, \citenamefont {Troyer}, \citenamefont {Dai},\ and\ \citenamefont
  {Bernevig}}]{soluyanov2015type}%
  \BibitemOpen
  \bibfield  {author} {\bibinfo {author} {\bibfnamefont {A.~A.}\ \bibnamefont
  {Soluyanov}}, \bibinfo {author} {\bibfnamefont {D.}~\bibnamefont {Gresch}},
  \bibinfo {author} {\bibfnamefont {Z.}~\bibnamefont {Wang}}, \bibinfo {author}
  {\bibfnamefont {Q.}~\bibnamefont {Wu}}, \bibinfo {author} {\bibfnamefont
  {M.}~\bibnamefont {Troyer}}, \bibinfo {author} {\bibfnamefont
  {X.}~\bibnamefont {Dai}}, \ and\ \bibinfo {author} {\bibfnamefont {B.~A.}\
  \bibnamefont {Bernevig}},\ }\href@noop {} {\bibfield  {journal} {\bibinfo
  {journal} {Nat.},\ }\textbf {\bibinfo {volume} {527}},\ \bibinfo {pages}
  {495} (\bibinfo {year} {2015})}\BibitemShut {NoStop}%
\bibitem [{\citenamefont {Yang}\ \emph {et~al.}(2015)\citenamefont {Yang},
  \citenamefont {Liu}, \citenamefont {Sun}, \citenamefont {Peng}, \citenamefont
  {Yang}, \citenamefont {Zhang}, \citenamefont {Zhou}, \citenamefont {Zhang},
  \citenamefont {Guo}, \citenamefont {Rahn} \emph {et~al.}}]{yang2015weyl}%
  \BibitemOpen
  \bibfield  {author} {\bibinfo {author} {\bibfnamefont {L.}~\bibnamefont
  {Yang}}, \bibinfo {author} {\bibfnamefont {Z.}~\bibnamefont {Liu}}, \bibinfo
  {author} {\bibfnamefont {Y.}~\bibnamefont {Sun}}, \bibinfo {author}
  {\bibfnamefont {H.}~\bibnamefont {Peng}}, \bibinfo {author} {\bibfnamefont
  {H.}~\bibnamefont {Yang}}, \bibinfo {author} {\bibfnamefont {T.}~\bibnamefont
  {Zhang}}, \bibinfo {author} {\bibfnamefont {B.}~\bibnamefont {Zhou}},
  \bibinfo {author} {\bibfnamefont {Y.}~\bibnamefont {Zhang}}, \bibinfo
  {author} {\bibfnamefont {Y.}~\bibnamefont {Guo}}, \bibinfo {author}
  {\bibfnamefont {M.}~\bibnamefont {Rahn}},  \emph {et~al.},\ }\href@noop {}
  {\bibfield  {journal} {\bibinfo  {journal} {Nat. phys.},\ }\textbf {\bibinfo
  {volume} {11}},\ \bibinfo {pages} {728} (\bibinfo {year} {2015})}\BibitemShut
  {NoStop}%
\bibitem [{\citenamefont {Lv}\ \emph {et~al.}(2015)\citenamefont {Lv},
  \citenamefont {Weng}, \citenamefont {Fu}, \citenamefont {Wang}, \citenamefont
  {Miao}, \citenamefont {Ma}, \citenamefont {Richard}, \citenamefont {Huang},
  \citenamefont {Zhao}, \citenamefont {Chen} \emph
  {et~al.}}]{lv2015experimental}%
  \BibitemOpen
  \bibfield  {author} {\bibinfo {author} {\bibfnamefont {B.}~\bibnamefont
  {Lv}}, \bibinfo {author} {\bibfnamefont {H.}~\bibnamefont {Weng}}, \bibinfo
  {author} {\bibfnamefont {B.}~\bibnamefont {Fu}}, \bibinfo {author}
  {\bibfnamefont {X.~P.}\ \bibnamefont {Wang}}, \bibinfo {author}
  {\bibfnamefont {H.}~\bibnamefont {Miao}}, \bibinfo {author} {\bibfnamefont
  {J.}~\bibnamefont {Ma}}, \bibinfo {author} {\bibfnamefont {P.}~\bibnamefont
  {Richard}}, \bibinfo {author} {\bibfnamefont {X.}~\bibnamefont {Huang}},
  \bibinfo {author} {\bibfnamefont {L.}~\bibnamefont {Zhao}}, \bibinfo {author}
  {\bibfnamefont {G.}~\bibnamefont {Chen}},  \emph {et~al.},\ }\href@noop {}
  {\bibfield  {journal} {\bibinfo  {journal} {Phys. Rev. X.},\ }\textbf
  {\bibinfo {volume} {5}},\ \bibinfo {pages} {031013} (\bibinfo {year}
  {2015})}\BibitemShut {NoStop}%
\bibitem [{\citenamefont {Xu}\ \emph {et~al.}(2011)\citenamefont {Xu},
  \citenamefont {Weng}, \citenamefont {Wang}, \citenamefont {Dai},\ and\
  \citenamefont {Fang}}]{xu2011chern}%
  \BibitemOpen
  \bibfield  {author} {\bibinfo {author} {\bibfnamefont {G.}~\bibnamefont
  {Xu}}, \bibinfo {author} {\bibfnamefont {H.}~\bibnamefont {Weng}}, \bibinfo
  {author} {\bibfnamefont {Z.}~\bibnamefont {Wang}}, \bibinfo {author}
  {\bibfnamefont {X.}~\bibnamefont {Dai}}, \ and\ \bibinfo {author}
  {\bibfnamefont {Z.}~\bibnamefont {Fang}},\ }\href@noop {} {\bibfield
  {journal} {\bibinfo  {journal} {Phys. Rev. Lett.},\ }\textbf {\bibinfo
  {volume} {107}},\ \bibinfo {pages} {186806} (\bibinfo {year}
  {2011})}\BibitemShut {NoStop}%
\bibitem [{\citenamefont {Armitage}\ \emph {et~al.}(2018)\citenamefont
  {Armitage}, \citenamefont {Mele},\ and\ \citenamefont
  {Vishwanath}}]{armitage2018weyl}%
  \BibitemOpen
  \bibfield  {author} {\bibinfo {author} {\bibfnamefont {N.}~\bibnamefont
  {Armitage}}, \bibinfo {author} {\bibfnamefont {E.}~\bibnamefont {Mele}}, \
  and\ \bibinfo {author} {\bibfnamefont {A.}~\bibnamefont {Vishwanath}},\
  }\href@noop {} {\bibfield  {journal} {\bibinfo  {journal} {Rev. Mod. Phys.},\
  }\textbf {\bibinfo {volume} {90}},\ \bibinfo {pages} {015001} (\bibinfo
  {year} {2018})}\BibitemShut {NoStop}%
\bibitem [{\citenamefont {Wang}\ \emph {et~al.}(2016)\citenamefont {Wang},
  \citenamefont {Vergniory}, \citenamefont {Kushwaha}, \citenamefont
  {Hirschberger}, \citenamefont {Chulkov}, \citenamefont {Ernst}, \citenamefont
  {Ong}, \citenamefont {Cava},\ and\ \citenamefont {Bernevig}}]{wang2016time}%
  \BibitemOpen
  \bibfield  {author} {\bibinfo {author} {\bibfnamefont {Z.}~\bibnamefont
  {Wang}}, \bibinfo {author} {\bibfnamefont {M.}~\bibnamefont {Vergniory}},
  \bibinfo {author} {\bibfnamefont {S.}~\bibnamefont {Kushwaha}}, \bibinfo
  {author} {\bibfnamefont {M.}~\bibnamefont {Hirschberger}}, \bibinfo {author}
  {\bibfnamefont {E.}~\bibnamefont {Chulkov}}, \bibinfo {author} {\bibfnamefont
  {A.}~\bibnamefont {Ernst}}, \bibinfo {author} {\bibfnamefont {N.~P.}\
  \bibnamefont {Ong}}, \bibinfo {author} {\bibfnamefont {R.~J.}\ \bibnamefont
  {Cava}}, \ and\ \bibinfo {author} {\bibfnamefont {B.~A.}\ \bibnamefont
  {Bernevig}},\ }\href@noop {} {\bibfield  {journal} {\bibinfo  {journal}
  {Phys. Rev. Lett.},\ }\textbf {\bibinfo {volume} {117}},\ \bibinfo {pages}
  {236401} (\bibinfo {year} {2016})}\BibitemShut {NoStop}%
\bibitem [{\citenamefont {Tokura}\ and\ \citenamefont
  {Kanazawa}(2020)}]{tokura2020magnetic}%
  \BibitemOpen
  \bibfield  {author} {\bibinfo {author} {\bibfnamefont {Y.}~\bibnamefont
  {Tokura}}\ and\ \bibinfo {author} {\bibfnamefont {N.}~\bibnamefont
  {Kanazawa}},\ }\href@noop {} {\bibfield  {journal} {\bibinfo  {journal}
  {Chem. Rev.},\ }\textbf {\bibinfo {volume} {121}},\ \bibinfo {pages} {2857}
  (\bibinfo {year} {2020})}\BibitemShut {NoStop}%
\bibitem [{\citenamefont {M.~Hirschberger}(2016)}]{GdPtBinature}%
  \BibitemOpen
  \bibfield  {author} {\bibinfo {author} {\bibfnamefont {Z.~W. Q. G. S. L. C.
  A. B. B. A. B. R. J. C. N. P.~O.}\ \bibnamefont {M.~Hirschberger},
  \bibfnamefont {S.~Kushwaha}},\ }\href@noop {} {\bibfield  {journal} {\bibinfo
   {journal} {Nat. Mater.},\ }\textbf {\bibinfo {volume} {15}},\ \bibinfo
  {pages} {1161} (\bibinfo {year} {2016})}\BibitemShut {NoStop}%
\bibitem [{\citenamefont {Bhattacharya}\ \emph {et~al.}(2024)\citenamefont
  {Bhattacharya}, \citenamefont {PC}, \citenamefont {Ahmed}, \citenamefont
  {Kurebayashi}, \citenamefont {Tretiakov}, \citenamefont {Satpati},
  \citenamefont {DuttaGupta}, \citenamefont {Alam},\ and\ \citenamefont
  {Das}}]{bhattacharya2024giant}%
  \BibitemOpen
  \bibfield  {author} {\bibinfo {author} {\bibfnamefont {A.}~\bibnamefont
  {Bhattacharya}}, \bibinfo {author} {\bibfnamefont {S.}~\bibnamefont {PC}},
  \bibinfo {author} {\bibfnamefont {A.}~\bibnamefont {Ahmed}}, \bibinfo
  {author} {\bibfnamefont {D.}~\bibnamefont {Kurebayashi}}, \bibinfo {author}
  {\bibfnamefont {O.~A.}\ \bibnamefont {Tretiakov}}, \bibinfo {author}
  {\bibfnamefont {B.}~\bibnamefont {Satpati}}, \bibinfo {author} {\bibfnamefont
  {S.}~\bibnamefont {DuttaGupta}}, \bibinfo {author} {\bibfnamefont
  {A.}~\bibnamefont {Alam}}, \ and\ \bibinfo {author} {\bibfnamefont
  {I.}~\bibnamefont {Das}},\ }\href@noop {} {\bibfield  {journal} {\bibinfo
  {journal} {Adv. Funct. Mater.},\ \bibinfo {pages} {2424841}} (\bibinfo {year}
  {2024})}\BibitemShut {NoStop}%
\bibitem [{\citenamefont {Chang}\ \emph {et~al.}(2018)\citenamefont {Chang},
  \citenamefont {Singh}, \citenamefont {Xu}, \citenamefont {Bian},
  \citenamefont {Huang}, \citenamefont {Hsu}, \citenamefont {Belopolski},
  \citenamefont {Alidoust}, \citenamefont {Sanchez}, \citenamefont {Zheng}
  \emph {et~al.}}]{chang2018magnetic}%
  \BibitemOpen
  \bibfield  {author} {\bibinfo {author} {\bibfnamefont {G.}~\bibnamefont
  {Chang}}, \bibinfo {author} {\bibfnamefont {B.}~\bibnamefont {Singh}},
  \bibinfo {author} {\bibfnamefont {S.-Y.}\ \bibnamefont {Xu}}, \bibinfo
  {author} {\bibfnamefont {G.}~\bibnamefont {Bian}}, \bibinfo {author}
  {\bibfnamefont {S.-M.}\ \bibnamefont {Huang}}, \bibinfo {author}
  {\bibfnamefont {C.-H.}\ \bibnamefont {Hsu}}, \bibinfo {author} {\bibfnamefont
  {I.}~\bibnamefont {Belopolski}}, \bibinfo {author} {\bibfnamefont
  {N.}~\bibnamefont {Alidoust}}, \bibinfo {author} {\bibfnamefont {D.~S.}\
  \bibnamefont {Sanchez}}, \bibinfo {author} {\bibfnamefont {H.}~\bibnamefont
  {Zheng}},  \emph {et~al.},\ }\href@noop {} {\bibfield  {journal} {\bibinfo
  {journal} {Phys. Rev. B},\ }\textbf {\bibinfo {volume} {97}},\ \bibinfo
  {pages} {041104} (\bibinfo {year} {2018})}\BibitemShut {NoStop}%
\bibitem [{\citenamefont {Li}\ \emph {et~al.}(2023)\citenamefont {Li},
  \citenamefont {Zhang}, \citenamefont {Wang}, \citenamefont {Liu},
  \citenamefont {Guo}, \citenamefont {Rienks}, \citenamefont {Chen},
  \citenamefont {Bertran}, \citenamefont {Yang}, \citenamefont {Phuyal} \emph
  {et~al.}}]{li2023emergence}%
  \BibitemOpen
  \bibfield  {author} {\bibinfo {author} {\bibfnamefont {C.}~\bibnamefont
  {Li}}, \bibinfo {author} {\bibfnamefont {J.}~\bibnamefont {Zhang}}, \bibinfo
  {author} {\bibfnamefont {Y.}~\bibnamefont {Wang}}, \bibinfo {author}
  {\bibfnamefont {H.}~\bibnamefont {Liu}}, \bibinfo {author} {\bibfnamefont
  {Q.}~\bibnamefont {Guo}}, \bibinfo {author} {\bibfnamefont {E.}~\bibnamefont
  {Rienks}}, \bibinfo {author} {\bibfnamefont {W.}~\bibnamefont {Chen}},
  \bibinfo {author} {\bibfnamefont {F.}~\bibnamefont {Bertran}}, \bibinfo
  {author} {\bibfnamefont {H.}~\bibnamefont {Yang}}, \bibinfo {author}
  {\bibfnamefont {D.}~\bibnamefont {Phuyal}},  \emph {et~al.},\ }\href@noop {}
  {\bibfield  {journal} {\bibinfo  {journal} {Nat. Commun.},\ }\textbf
  {\bibinfo {volume} {14}},\ \bibinfo {pages} {7185} (\bibinfo {year}
  {2023})}\BibitemShut {NoStop}%
\bibitem [{\citenamefont {Hohenberg}\ and\ \citenamefont
  {Kohn}(1964)}]{PhysRev.136.B864}%
  \BibitemOpen
  \bibfield  {author} {\bibinfo {author} {\bibfnamefont {P.}~\bibnamefont
  {Hohenberg}}\ and\ \bibinfo {author} {\bibfnamefont {W.}~\bibnamefont
  {Kohn}},\ }\href@noop {} {\bibfield  {journal} {\bibinfo  {journal} {Phys.
  Rev.},\ }\textbf {\bibinfo {volume} {136}},\ \bibinfo {pages} {B864}
  (\bibinfo {year} {1964})}\BibitemShut {NoStop}%
\bibitem [{\citenamefont {Kresse}\ and\ \citenamefont
  {Joubert}(1999)}]{kresse1999ultrasoft}%
  \BibitemOpen
  \bibfield  {author} {\bibinfo {author} {\bibfnamefont {G.}~\bibnamefont
  {Kresse}}\ and\ \bibinfo {author} {\bibfnamefont {D.}~\bibnamefont
  {Joubert}},\ }\href@noop {} {\bibfield  {journal} {\bibinfo  {journal} {Phys.
  Rev. B},\ }\textbf {\bibinfo {volume} {59}},\ \bibinfo {pages} {1758}
  (\bibinfo {year} {1999})}\BibitemShut {NoStop}%
\bibitem [{\citenamefont {Perdew}\ \emph {et~al.}(1996)\citenamefont {Perdew},
  \citenamefont {Burke},\ and\ \citenamefont
  {Ernzerhof}}]{perdew1996generalized}%
  \BibitemOpen
  \bibfield  {author} {\bibinfo {author} {\bibfnamefont {J.~P.}\ \bibnamefont
  {Perdew}}, \bibinfo {author} {\bibfnamefont {K.}~\bibnamefont {Burke}}, \
  and\ \bibinfo {author} {\bibfnamefont {M.}~\bibnamefont {Ernzerhof}},\
  }\href@noop {} {\bibfield  {journal} {\bibinfo  {journal} {Phys. Rev.
  Lett.},\ }\textbf {\bibinfo {volume} {77}},\ \bibinfo {pages} {3865}
  (\bibinfo {year} {1996})}\BibitemShut {NoStop}%
\bibitem [{SM(2025)}]{SM}%
  \BibitemOpen
  \href@noop {} {\bibfield  {journal} {\bibinfo  {journal} {Supplementary URL
  to be added by journal}} (\bibinfo {year} {2025})}\BibitemShut {NoStop}%
\bibitem [{\citenamefont {Canepa}\ \emph
  {et~al.}(2000){\natexlab{a}}\citenamefont {Canepa}, \citenamefont {Cirafici},
  \citenamefont {Fornasini}, \citenamefont {Manfrinetti}, \citenamefont
  {Merlo}, \citenamefont {Palenzona},\ and\ \citenamefont
  {Pani}}]{canepa2000crystal}%
  \BibitemOpen
  \bibfield  {author} {\bibinfo {author} {\bibfnamefont {F.}~\bibnamefont
  {Canepa}}, \bibinfo {author} {\bibfnamefont {S.}~\bibnamefont {Cirafici}},
  \bibinfo {author} {\bibfnamefont {M.}~\bibnamefont {Fornasini}}, \bibinfo
  {author} {\bibfnamefont {P.}~\bibnamefont {Manfrinetti}}, \bibinfo {author}
  {\bibfnamefont {F.}~\bibnamefont {Merlo}}, \bibinfo {author} {\bibfnamefont
  {A.}~\bibnamefont {Palenzona}}, \ and\ \bibinfo {author} {\bibfnamefont
  {M.}~\bibnamefont {Pani}},\ }\href@noop {} {\bibfield  {journal} {\bibinfo
  {journal} {J. Alloys Compd.},\ }\textbf {\bibinfo {volume} {297}},\ \bibinfo
  {pages} {109} (\bibinfo {year} {2000}{\natexlab{a}})}\BibitemShut {NoStop}%
\bibitem [{\citenamefont {Canepa}\ \emph {et~al.}(2005)\citenamefont {Canepa},
  \citenamefont {Napoletano}, \citenamefont {Palenzona}, \citenamefont {Moze},\
  and\ \citenamefont {Kockelmann}}]{canepa2004ferromagnetic}%
  \BibitemOpen
  \bibfield  {author} {\bibinfo {author} {\bibfnamefont {F.}~\bibnamefont
  {Canepa}}, \bibinfo {author} {\bibfnamefont {M.}~\bibnamefont {Napoletano}},
  \bibinfo {author} {\bibfnamefont {A.}~\bibnamefont {Palenzona}}, \bibinfo
  {author} {\bibfnamefont {O.}~\bibnamefont {Moze}}, \ and\ \bibinfo {author}
  {\bibfnamefont {W.}~\bibnamefont {Kockelmann}},\ }\href@noop {} {\bibfield
  {journal} {\bibinfo  {journal} {J. Phys. Condens. Matter},\ }\textbf
  {\bibinfo {volume} {17}},\ \bibinfo {pages} {373} (\bibinfo {year}
  {2005})}\BibitemShut {NoStop}%
\bibitem [{\citenamefont {Canepa}\ \emph
  {et~al.}(2000){\natexlab{b}}\citenamefont {Canepa}, \citenamefont
  {Napoletano}, \citenamefont {Manfrinetti}, \citenamefont {Palenzona},
  \citenamefont {Cirafici},\ and\ \citenamefont {Merlo}}]{canepa2000magnetic}%
  \BibitemOpen
  \bibfield  {author} {\bibinfo {author} {\bibfnamefont {F.}~\bibnamefont
  {Canepa}}, \bibinfo {author} {\bibfnamefont {M.}~\bibnamefont {Napoletano}},
  \bibinfo {author} {\bibfnamefont {P.}~\bibnamefont {Manfrinetti}}, \bibinfo
  {author} {\bibfnamefont {A.}~\bibnamefont {Palenzona}}, \bibinfo {author}
  {\bibfnamefont {S.}~\bibnamefont {Cirafici}}, \ and\ \bibinfo {author}
  {\bibfnamefont {F.}~\bibnamefont {Merlo}},\ }\href@noop {} {\bibfield
  {journal} {\bibinfo  {journal} {J. Magn. Magn. Mater.},\ }\textbf {\bibinfo
  {volume} {220}},\ \bibinfo {pages} {39} (\bibinfo {year}
  {2000}{\natexlab{b}})}\BibitemShut {NoStop}%
\bibitem [{\citenamefont {Takagi}\ \emph {et~al.}(2018)\citenamefont {Takagi},
  \citenamefont {White}, \citenamefont {Hayami}, \citenamefont {Arita},
  \citenamefont {Honecker}, \citenamefont {R{\o}nnow}, \citenamefont {Tokura},\
  and\ \citenamefont {Seki}}]{takagi2018multiple}%
  \BibitemOpen
  \bibfield  {author} {\bibinfo {author} {\bibfnamefont {R.}~\bibnamefont
  {Takagi}}, \bibinfo {author} {\bibfnamefont {J.}~\bibnamefont {White}},
  \bibinfo {author} {\bibfnamefont {S.}~\bibnamefont {Hayami}}, \bibinfo
  {author} {\bibfnamefont {R.}~\bibnamefont {Arita}}, \bibinfo {author}
  {\bibfnamefont {D.}~\bibnamefont {Honecker}}, \bibinfo {author}
  {\bibfnamefont {H.}~\bibnamefont {R{\o}nnow}}, \bibinfo {author}
  {\bibfnamefont {Y.}~\bibnamefont {Tokura}}, \ and\ \bibinfo {author}
  {\bibfnamefont {S.}~\bibnamefont {Seki}},\ }\href@noop {} {\bibfield
  {journal} {\bibinfo  {journal} {Sci. adv.},\ }\textbf {\bibinfo {volume}
  {4}},\ \bibinfo {pages} {eaau3402} (\bibinfo {year} {2018})}\BibitemShut
  {NoStop}%
\bibitem [{\citenamefont {Madduri}\ \emph {et~al.}(2020)\citenamefont
  {Madduri}, \citenamefont {Sen}, \citenamefont {Giri}, \citenamefont
  {Chakrabartty}, \citenamefont {Manna}, \citenamefont {Parkin},\ and\
  \citenamefont {Nayak}}]{madduri2020ac}%
  \BibitemOpen
  \bibfield  {author} {\bibinfo {author} {\bibfnamefont {P.~P.}\ \bibnamefont
  {Madduri}}, \bibinfo {author} {\bibfnamefont {S.}~\bibnamefont {Sen}},
  \bibinfo {author} {\bibfnamefont {B.}~\bibnamefont {Giri}}, \bibinfo {author}
  {\bibfnamefont {D.}~\bibnamefont {Chakrabartty}}, \bibinfo {author}
  {\bibfnamefont {S.~K.}\ \bibnamefont {Manna}}, \bibinfo {author}
  {\bibfnamefont {S.~S.}\ \bibnamefont {Parkin}}, \ and\ \bibinfo {author}
  {\bibfnamefont {A.~K.}\ \bibnamefont {Nayak}},\ }\href@noop {} {\bibfield
  {journal} {\bibinfo  {journal} {Phys. Rev. B},\ }\textbf {\bibinfo {volume}
  {102}},\ \bibinfo {pages} {174402} (\bibinfo {year} {2020})}\BibitemShut
  {NoStop}%
\bibitem [{\citenamefont {Sen}\ \emph {et~al.}(2019)\citenamefont {Sen},
  \citenamefont {Singh}, \citenamefont {Mukharjee}, \citenamefont {Nath},\ and\
  \citenamefont {Nayak}}]{sen2019observation}%
  \BibitemOpen
  \bibfield  {author} {\bibinfo {author} {\bibfnamefont {S.}~\bibnamefont
  {Sen}}, \bibinfo {author} {\bibfnamefont {C.}~\bibnamefont {Singh}}, \bibinfo
  {author} {\bibfnamefont {P.~K.}\ \bibnamefont {Mukharjee}}, \bibinfo {author}
  {\bibfnamefont {R.}~\bibnamefont {Nath}}, \ and\ \bibinfo {author}
  {\bibfnamefont {A.~K.}\ \bibnamefont {Nayak}},\ }\href@noop {} {\bibfield
  {journal} {\bibinfo  {journal} {Phys. Rev. B},\ }\textbf {\bibinfo {volume}
  {99}},\ \bibinfo {pages} {134404} (\bibinfo {year} {2019})}\BibitemShut
  {NoStop}%
\bibitem [{\citenamefont {Bannenberg}\ \emph {et~al.}(2018)\citenamefont
  {Bannenberg}, \citenamefont {Weber}, \citenamefont {Lefering}, \citenamefont
  {Wolf},\ and\ \citenamefont {Pappas}}]{p31}%
  \BibitemOpen
  \bibfield  {author} {\bibinfo {author} {\bibfnamefont {L.~J.}\ \bibnamefont
  {Bannenberg}}, \bibinfo {author} {\bibfnamefont {F.}~\bibnamefont {Weber}},
  \bibinfo {author} {\bibfnamefont {A.~J.~E.}\ \bibnamefont {Lefering}},
  \bibinfo {author} {\bibfnamefont {T.}~\bibnamefont {Wolf}}, \ and\ \bibinfo
  {author} {\bibfnamefont {C.}~\bibnamefont {Pappas}},\ }\href@noop {}
  {\bibfield  {journal} {\bibinfo  {journal} {Phys. Rev. B},\ }\textbf
  {\bibinfo {volume} {98}},\ \bibinfo {pages} {184430} (\bibinfo {year}
  {2018})}\BibitemShut {NoStop}%
\bibitem [{\citenamefont {Bannenberg}\ \emph {et~al.}(2016)\citenamefont
  {Bannenberg}, \citenamefont {Lefering}, \citenamefont {Kakurai},
  \citenamefont {Onose}, \citenamefont {Endoh}, \citenamefont {Tokura},\ and\
  \citenamefont {Pappas}}]{p30}%
  \BibitemOpen
  \bibfield  {author} {\bibinfo {author} {\bibfnamefont {L.~J.}\ \bibnamefont
  {Bannenberg}}, \bibinfo {author} {\bibfnamefont {A.~J.~E.}\ \bibnamefont
  {Lefering}}, \bibinfo {author} {\bibfnamefont {K.}~\bibnamefont {Kakurai}},
  \bibinfo {author} {\bibfnamefont {Y.}~\bibnamefont {Onose}}, \bibinfo
  {author} {\bibfnamefont {Y.}~\bibnamefont {Endoh}}, \bibinfo {author}
  {\bibfnamefont {Y.}~\bibnamefont {Tokura}}, \ and\ \bibinfo {author}
  {\bibfnamefont {C.}~\bibnamefont {Pappas}},\ }\href@noop {} {\bibfield
  {journal} {\bibinfo  {journal} {Phys. Rev. B},\ }\textbf {\bibinfo {volume}
  {94}},\ \bibinfo {pages} {134433} (\bibinfo {year} {2016})}\BibitemShut
  {NoStop}%
\bibitem [{\citenamefont {Wilhelm}\ \emph {et~al.}(2011)\citenamefont
  {Wilhelm}, \citenamefont {Baenitz}, \citenamefont {Schmidt}, \citenamefont
  {R\"o\ss{}ler}, \citenamefont {Leonov},\ and\ \citenamefont
  {Bogdanov}}]{p29}%
  \BibitemOpen
  \bibfield  {author} {\bibinfo {author} {\bibfnamefont {H.}~\bibnamefont
  {Wilhelm}}, \bibinfo {author} {\bibfnamefont {M.}~\bibnamefont {Baenitz}},
  \bibinfo {author} {\bibfnamefont {M.}~\bibnamefont {Schmidt}}, \bibinfo
  {author} {\bibfnamefont {U.~K.}\ \bibnamefont {R\"o\ss{}ler}}, \bibinfo
  {author} {\bibfnamefont {A.~A.}\ \bibnamefont {Leonov}}, \ and\ \bibinfo
  {author} {\bibfnamefont {A.~N.}\ \bibnamefont {Bogdanov}},\ }\href@noop {}
  {\bibfield  {journal} {\bibinfo  {journal} {Phys. Rev. Lett.},\ }\textbf
  {\bibinfo {volume} {107}},\ \bibinfo {pages} {127203} (\bibinfo {year}
  {2011})}\BibitemShut {NoStop}%
\bibitem [{\citenamefont {Bauer}\ and\ \citenamefont {Pfleiderer}(2012)}]{p28}%
  \BibitemOpen
  \bibfield  {author} {\bibinfo {author} {\bibfnamefont {A.}~\bibnamefont
  {Bauer}}\ and\ \bibinfo {author} {\bibfnamefont {C.}~\bibnamefont
  {Pfleiderer}},\ }\href@noop {} {\bibfield  {journal} {\bibinfo  {journal}
  {Phys. R. B},\ }\textbf {\bibinfo {volume} {85}},\ \bibinfo {pages} {214418}
  (\bibinfo {year} {2012})}\BibitemShut {NoStop}%
\bibitem [{\citenamefont {Kurumaji}\ \emph {et~al.}(2019)\citenamefont
  {Kurumaji}, \citenamefont {Nakajima}, \citenamefont {Hirschberger},
  \citenamefont {Kikkawa}, \citenamefont {Yamasaki}, \citenamefont {Sagayama},
  \citenamefont {Nakao}, \citenamefont {Taguchi}, \citenamefont {Arima},\ and\
  \citenamefont {Tokura}}]{p34}%
  \BibitemOpen
  \bibfield  {author} {\bibinfo {author} {\bibfnamefont {T.}~\bibnamefont
  {Kurumaji}}, \bibinfo {author} {\bibfnamefont {T.}~\bibnamefont {Nakajima}},
  \bibinfo {author} {\bibfnamefont {M.}~\bibnamefont {Hirschberger}}, \bibinfo
  {author} {\bibfnamefont {A.}~\bibnamefont {Kikkawa}}, \bibinfo {author}
  {\bibfnamefont {Y.}~\bibnamefont {Yamasaki}}, \bibinfo {author}
  {\bibfnamefont {H.}~\bibnamefont {Sagayama}}, \bibinfo {author}
  {\bibfnamefont {H.}~\bibnamefont {Nakao}}, \bibinfo {author} {\bibfnamefont
  {Y.}~\bibnamefont {Taguchi}}, \bibinfo {author} {\bibfnamefont {T.-h.}\
  \bibnamefont {Arima}}, \ and\ \bibinfo {author} {\bibfnamefont
  {Y.}~\bibnamefont {Tokura}},\ }\href@noop {} {\bibfield  {journal} {\bibinfo
  {journal} {Sci.},\ }\textbf {\bibinfo {volume} {365}},\ \bibinfo {pages}
  {914} (\bibinfo {year} {2019})}\BibitemShut {NoStop}%
\bibitem [{\citenamefont {Rawat}\ and\ \citenamefont
  {Das}(2001)}]{rawat2001magnetic}%
  \BibitemOpen
  \bibfield  {author} {\bibinfo {author} {\bibfnamefont {R.}~\bibnamefont
  {Rawat}}\ and\ \bibinfo {author} {\bibfnamefont {I.}~\bibnamefont {Das}},\
  }\href@noop {} {\bibfield  {journal} {\bibinfo  {journal} {J. Magn. Magn.
  Mater.},\ }\textbf {\bibinfo {volume} {236}},\ \bibinfo {pages} {285}
  (\bibinfo {year} {2001})}\BibitemShut {NoStop}%
\bibitem [{\citenamefont {Mishra}\ \emph {et~al.}(2024)\citenamefont {Mishra},
  \citenamefont {Kengle}, \citenamefont {Thompson}, \citenamefont {Scheie},
  \citenamefont {Thomas}, \citenamefont {Ronning}, \citenamefont {Rosa} \emph
  {et~al.}}]{mishra2024evidence}%
  \BibitemOpen
  \bibfield  {author} {\bibinfo {author} {\bibfnamefont {S.}~\bibnamefont
  {Mishra}}, \bibinfo {author} {\bibfnamefont {C.~S.}\ \bibnamefont {Kengle}},
  \bibinfo {author} {\bibfnamefont {J.~D.}\ \bibnamefont {Thompson}}, \bibinfo
  {author} {\bibfnamefont {A.~O.}\ \bibnamefont {Scheie}}, \bibinfo {author}
  {\bibfnamefont {S.}~\bibnamefont {Thomas}}, \bibinfo {author} {\bibfnamefont
  {F.}~\bibnamefont {Ronning}}, \bibinfo {author} {\bibfnamefont {P.~F.}\
  \bibnamefont {Rosa}},  \emph {et~al.},\ }\href@noop {} {\bibfield  {journal}
  {\bibinfo  {journal} {arXiv:2412.10998}} (\bibinfo {year}
  {2024})}\BibitemShut {NoStop}%
\bibitem [{\citenamefont {Zhang}\ \emph {et~al.}(2020)\citenamefont {Zhang},
  \citenamefont {Huang}, \citenamefont {Hao}, \citenamefont {Yang},
  \citenamefont {Noordhoek}, \citenamefont {Pandey}, \citenamefont {Zhou},\
  and\ \citenamefont {Liu}}]{zhang2020anomalous}%
  \BibitemOpen
  \bibfield  {author} {\bibinfo {author} {\bibfnamefont {H.}~\bibnamefont
  {Zhang}}, \bibinfo {author} {\bibfnamefont {Q.}~\bibnamefont {Huang}},
  \bibinfo {author} {\bibfnamefont {L.}~\bibnamefont {Hao}}, \bibinfo {author}
  {\bibfnamefont {J.}~\bibnamefont {Yang}}, \bibinfo {author} {\bibfnamefont
  {K.}~\bibnamefont {Noordhoek}}, \bibinfo {author} {\bibfnamefont
  {S.}~\bibnamefont {Pandey}}, \bibinfo {author} {\bibfnamefont
  {H.}~\bibnamefont {Zhou}}, \ and\ \bibinfo {author} {\bibfnamefont
  {J.}~\bibnamefont {Liu}},\ }\href@noop {} {\bibfield  {journal} {\bibinfo
  {journal} {New J. Phys.},\ }\textbf {\bibinfo {volume} {22}},\ \bibinfo
  {pages} {083056} (\bibinfo {year} {2020})}\BibitemShut {NoStop}%
\bibitem [{\citenamefont {Saha}\ \emph {et~al.}(1999)\citenamefont {Saha},
  \citenamefont {Sugawara}, \citenamefont {Matsuda}, \citenamefont {Sato},
  \citenamefont {Mallik},\ and\ \citenamefont
  {Sampathkumaran}}]{saha1999magnetic}%
  \BibitemOpen
  \bibfield  {author} {\bibinfo {author} {\bibfnamefont {S.}~\bibnamefont
  {Saha}}, \bibinfo {author} {\bibfnamefont {H.}~\bibnamefont {Sugawara}},
  \bibinfo {author} {\bibfnamefont {T.}~\bibnamefont {Matsuda}}, \bibinfo
  {author} {\bibfnamefont {H.}~\bibnamefont {Sato}}, \bibinfo {author}
  {\bibfnamefont {R.}~\bibnamefont {Mallik}}, \ and\ \bibinfo {author}
  {\bibfnamefont {E.}~\bibnamefont {Sampathkumaran}},\ }\href@noop {}
  {\bibfield  {journal} {\bibinfo  {journal} {Phys. Rev. B},\ }\textbf
  {\bibinfo {volume} {60}},\ \bibinfo {pages} {12162} (\bibinfo {year}
  {1999})}\BibitemShut {NoStop}%
\bibitem [{\citenamefont {You}\ \emph {et~al.}(2019)\citenamefont {You},
  \citenamefont {Gong}, \citenamefont {Li}, \citenamefont {Li}, \citenamefont
  {Zhu}, \citenamefont {Tang}, \citenamefont {Liu}, \citenamefont {Yao},
  \citenamefont {Xu}, \citenamefont {Xu} \emph {et~al.}}]{p35}%
  \BibitemOpen
  \bibfield  {author} {\bibinfo {author} {\bibfnamefont {Y.}~\bibnamefont
  {You}}, \bibinfo {author} {\bibfnamefont {Y.}~\bibnamefont {Gong}}, \bibinfo
  {author} {\bibfnamefont {H.}~\bibnamefont {Li}}, \bibinfo {author}
  {\bibfnamefont {Z.}~\bibnamefont {Li}}, \bibinfo {author} {\bibfnamefont
  {M.}~\bibnamefont {Zhu}}, \bibinfo {author} {\bibfnamefont {J.}~\bibnamefont
  {Tang}}, \bibinfo {author} {\bibfnamefont {E.}~\bibnamefont {Liu}}, \bibinfo
  {author} {\bibfnamefont {Y.}~\bibnamefont {Yao}}, \bibinfo {author}
  {\bibfnamefont {G.}~\bibnamefont {Xu}}, \bibinfo {author} {\bibfnamefont
  {F.}~\bibnamefont {Xu}},  \emph {et~al.},\ }\href@noop {} {\bibfield
  {journal} {\bibinfo  {journal} {Phys. Rev. B},\ }\textbf {\bibinfo {volume}
  {100}},\ \bibinfo {pages} {134441} (\bibinfo {year} {2019})}\BibitemShut
  {NoStop}%
\bibitem [{\citenamefont {Nagaosa}\ \emph {et~al.}(2010)\citenamefont
  {Nagaosa}, \citenamefont {Sinova}, \citenamefont {Onoda}, \citenamefont
  {MacDonald},\ and\ \citenamefont {Ong}}]{p11}%
  \BibitemOpen
  \bibfield  {author} {\bibinfo {author} {\bibfnamefont {N.}~\bibnamefont
  {Nagaosa}}, \bibinfo {author} {\bibfnamefont {J.}~\bibnamefont {Sinova}},
  \bibinfo {author} {\bibfnamefont {S.}~\bibnamefont {Onoda}}, \bibinfo
  {author} {\bibfnamefont {A.~H.}\ \bibnamefont {MacDonald}}, \ and\ \bibinfo
  {author} {\bibfnamefont {N.~P.}\ \bibnamefont {Ong}},\ }\href@noop {}
  {\bibfield  {journal} {\bibinfo  {journal} {Rev. Mod. Phys.},\ }\textbf
  {\bibinfo {volume} {82}},\ \bibinfo {pages} {1539} (\bibinfo {year}
  {2010})}\BibitemShut {NoStop}%
\bibitem [{\citenamefont {Tian}\ \emph {et~al.}(2009)\citenamefont {Tian},
  \citenamefont {Ye},\ and\ \citenamefont {Jin}}]{p13}%
  \BibitemOpen
  \bibfield  {author} {\bibinfo {author} {\bibfnamefont {Y.}~\bibnamefont
  {Tian}}, \bibinfo {author} {\bibfnamefont {L.}~\bibnamefont {Ye}}, \ and\
  \bibinfo {author} {\bibfnamefont {X.}~\bibnamefont {Jin}},\ }\href@noop {}
  {\bibfield  {journal} {\bibinfo  {journal} {Phys. Rev. Lett.},\ }\textbf
  {\bibinfo {volume} {103}},\ \bibinfo {pages} {087206} (\bibinfo {year}
  {2009})}\BibitemShut {NoStop}%
\bibitem [{\citenamefont {Saha}\ \emph {et~al.}(2023)\citenamefont {Saha},
  \citenamefont {Singh}, \citenamefont {Nagpal}, \citenamefont {Das},\ and\
  \citenamefont {Patnaik}}]{p14}%
  \BibitemOpen
  \bibfield  {author} {\bibinfo {author} {\bibfnamefont {P.}~\bibnamefont
  {Saha}}, \bibinfo {author} {\bibfnamefont {M.}~\bibnamefont {Singh}},
  \bibinfo {author} {\bibfnamefont {V.}~\bibnamefont {Nagpal}}, \bibinfo
  {author} {\bibfnamefont {P.}~\bibnamefont {Das}}, \ and\ \bibinfo {author}
  {\bibfnamefont {S.}~\bibnamefont {Patnaik}},\ }\href@noop {} {\bibfield
  {journal} {\bibinfo  {journal} {Phys. Rev. B},\ }\textbf {\bibinfo {volume}
  {107}},\ \bibinfo {pages} {035115} (\bibinfo {year} {2023})}\BibitemShut
  {NoStop}%
\bibitem [{\citenamefont {Shen}\ \emph
  {et~al.}(2020){\natexlab{a}}\citenamefont {Shen}, \citenamefont {Zeng},
  \citenamefont {Zhang}, \citenamefont {Sun}, \citenamefont {Yao},
  \citenamefont {Xi}, \citenamefont {Wang}, \citenamefont {Wu}, \citenamefont
  {Shen}, \citenamefont {Liu} \emph {et~al.}}]{p12}%
  \BibitemOpen
  \bibfield  {author} {\bibinfo {author} {\bibfnamefont {J.}~\bibnamefont
  {Shen}}, \bibinfo {author} {\bibfnamefont {Q.}~\bibnamefont {Zeng}}, \bibinfo
  {author} {\bibfnamefont {S.}~\bibnamefont {Zhang}}, \bibinfo {author}
  {\bibfnamefont {H.}~\bibnamefont {Sun}}, \bibinfo {author} {\bibfnamefont
  {Q.}~\bibnamefont {Yao}}, \bibinfo {author} {\bibfnamefont {X.}~\bibnamefont
  {Xi}}, \bibinfo {author} {\bibfnamefont {W.}~\bibnamefont {Wang}}, \bibinfo
  {author} {\bibfnamefont {G.}~\bibnamefont {Wu}}, \bibinfo {author}
  {\bibfnamefont {B.}~\bibnamefont {Shen}}, \bibinfo {author} {\bibfnamefont
  {Q.}~\bibnamefont {Liu}},  \emph {et~al.},\ }\href@noop {} {\bibfield
  {journal} {\bibinfo  {journal} {Adv. Funct. Mater.},\ }\textbf {\bibinfo
  {volume} {30}},\ \bibinfo {pages} {2000830} (\bibinfo {year}
  {2020}{\natexlab{a}})}\BibitemShut {NoStop}%
\bibitem [{\citenamefont {Shen}\ \emph
  {et~al.}(2020){\natexlab{b}}\citenamefont {Shen}, \citenamefont {Yao},
  \citenamefont {Zeng}, \citenamefont {Sun}, \citenamefont {Xi}, \citenamefont
  {Wu}, \citenamefont {Wang}, \citenamefont {Shen}, \citenamefont {Liu},\ and\
  \citenamefont {Liu}}]{p17}%
  \BibitemOpen
  \bibfield  {author} {\bibinfo {author} {\bibfnamefont {J.}~\bibnamefont
  {Shen}}, \bibinfo {author} {\bibfnamefont {Q.}~\bibnamefont {Yao}}, \bibinfo
  {author} {\bibfnamefont {Q.}~\bibnamefont {Zeng}}, \bibinfo {author}
  {\bibfnamefont {H.}~\bibnamefont {Sun}}, \bibinfo {author} {\bibfnamefont
  {X.}~\bibnamefont {Xi}}, \bibinfo {author} {\bibfnamefont {G.}~\bibnamefont
  {Wu}}, \bibinfo {author} {\bibfnamefont {W.}~\bibnamefont {Wang}}, \bibinfo
  {author} {\bibfnamefont {B.}~\bibnamefont {Shen}}, \bibinfo {author}
  {\bibfnamefont {Q.}~\bibnamefont {Liu}}, \ and\ \bibinfo {author}
  {\bibfnamefont {E.}~\bibnamefont {Liu}},\ }\href@noop {} {\bibfield
  {journal} {\bibinfo  {journal} {Phys. Rev. Lett.},\ }\textbf {\bibinfo
  {volume} {125}},\ \bibinfo {pages} {086602} (\bibinfo {year}
  {2020}{\natexlab{b}})}\BibitemShut {NoStop}%
\bibitem [{\citenamefont {Chakraborty}\ \emph {et~al.}(2022)\citenamefont
  {Chakraborty}, \citenamefont {Samanta}, \citenamefont {Guin}, \citenamefont
  {Noky}, \citenamefont {Robredo}, \citenamefont {Prasad}, \citenamefont
  {Kuebler}, \citenamefont {Shekhar}, \citenamefont {Vergniory},\ and\
  \citenamefont {Felser}}]{chakraborty2022berry}%
  \BibitemOpen
  \bibfield  {author} {\bibinfo {author} {\bibfnamefont {T.}~\bibnamefont
  {Chakraborty}}, \bibinfo {author} {\bibfnamefont {K.}~\bibnamefont
  {Samanta}}, \bibinfo {author} {\bibfnamefont {S.~N.}\ \bibnamefont {Guin}},
  \bibinfo {author} {\bibfnamefont {J.}~\bibnamefont {Noky}}, \bibinfo {author}
  {\bibfnamefont {I.}~\bibnamefont {Robredo}}, \bibinfo {author} {\bibfnamefont
  {S.}~\bibnamefont {Prasad}}, \bibinfo {author} {\bibfnamefont
  {J.}~\bibnamefont {Kuebler}}, \bibinfo {author} {\bibfnamefont
  {C.}~\bibnamefont {Shekhar}}, \bibinfo {author} {\bibfnamefont {M.~G.}\
  \bibnamefont {Vergniory}}, \ and\ \bibinfo {author} {\bibfnamefont
  {C.}~\bibnamefont {Felser}},\ }\href@noop {} {\bibfield  {journal} {\bibinfo
  {journal} {Phys. Rev. B},\ }\textbf {\bibinfo {volume} {106}},\ \bibinfo
  {pages} {155141} (\bibinfo {year} {2022})}\BibitemShut {NoStop}%
\bibitem [{\citenamefont {Miyasato}\ \emph {et~al.}(2007)\citenamefont
  {Miyasato}, \citenamefont {Abe}, \citenamefont {Fujii}, \citenamefont
  {Asamitsu}, \citenamefont {Onoda}, \citenamefont {Onose}, \citenamefont
  {Nagaosa},\ and\ \citenamefont {Tokura}}]{miyasato2007crossover}%
  \BibitemOpen
  \bibfield  {author} {\bibinfo {author} {\bibfnamefont {T.}~\bibnamefont
  {Miyasato}}, \bibinfo {author} {\bibfnamefont {N.}~\bibnamefont {Abe}},
  \bibinfo {author} {\bibfnamefont {T.}~\bibnamefont {Fujii}}, \bibinfo
  {author} {\bibfnamefont {A.}~\bibnamefont {Asamitsu}}, \bibinfo {author}
  {\bibfnamefont {S.}~\bibnamefont {Onoda}}, \bibinfo {author} {\bibfnamefont
  {Y.}~\bibnamefont {Onose}}, \bibinfo {author} {\bibfnamefont
  {N.}~\bibnamefont {Nagaosa}}, \ and\ \bibinfo {author} {\bibfnamefont
  {Y.}~\bibnamefont {Tokura}},\ }\href@noop {} {\bibfield  {journal} {\bibinfo
  {journal} {Phys. Rev. Lett.},\ }\textbf {\bibinfo {volume} {99}},\ \bibinfo
  {pages} {086602} (\bibinfo {year} {2007})}\BibitemShut {NoStop}%
\bibitem [{\citenamefont {Onoda}\ \emph {et~al.}(2006)\citenamefont {Onoda},
  \citenamefont {Sugimoto},\ and\ \citenamefont
  {Nagaosa}}]{onoda2006intrinsic}%
  \BibitemOpen
  \bibfield  {author} {\bibinfo {author} {\bibfnamefont {S.}~\bibnamefont
  {Onoda}}, \bibinfo {author} {\bibfnamefont {N.}~\bibnamefont {Sugimoto}}, \
  and\ \bibinfo {author} {\bibfnamefont {N.}~\bibnamefont {Nagaosa}},\
  }\href@noop {} {\bibfield  {journal} {\bibinfo  {journal} {Phys. Rev.
  Lett.},\ }\textbf {\bibinfo {volume} {97}},\ \bibinfo {pages} {126602}
  (\bibinfo {year} {2006})}\BibitemShut {NoStop}%
\bibitem [{\citenamefont {Lee}\ \emph {et~al.}(2007)\citenamefont {Lee},
  \citenamefont {Onose}, \citenamefont {Tokura},\ and\ \citenamefont
  {Ong}}]{p24}%
  \BibitemOpen
  \bibfield  {author} {\bibinfo {author} {\bibfnamefont {M.}~\bibnamefont
  {Lee}}, \bibinfo {author} {\bibfnamefont {Y.}~\bibnamefont {Onose}}, \bibinfo
  {author} {\bibfnamefont {Y.}~\bibnamefont {Tokura}}, \ and\ \bibinfo {author}
  {\bibfnamefont {N.~P.}\ \bibnamefont {Ong}},\ }\href@noop {} {\bibfield
  {journal} {\bibinfo  {journal} {Phys. Rev. B},\ }\textbf {\bibinfo {volume}
  {75}},\ \bibinfo {pages} {172403} (\bibinfo {year} {2007})}\BibitemShut
  {NoStop}%
\bibitem [{\citenamefont {S{\"u}rgers}\ \emph {et~al.}(2014)\citenamefont
  {S{\"u}rgers}, \citenamefont {Fischer}, \citenamefont {Winkel},\ and\
  \citenamefont {L{\"o}hneysen}}]{p22}%
  \BibitemOpen
  \bibfield  {author} {\bibinfo {author} {\bibfnamefont {C.}~\bibnamefont
  {S{\"u}rgers}}, \bibinfo {author} {\bibfnamefont {G.}~\bibnamefont
  {Fischer}}, \bibinfo {author} {\bibfnamefont {P.}~\bibnamefont {Winkel}}, \
  and\ \bibinfo {author} {\bibfnamefont {H.~V.}\ \bibnamefont
  {L{\"o}hneysen}},\ }\href@noop {} {\bibfield  {journal} {\bibinfo  {journal}
  {Nat. Commun.},\ }\textbf {\bibinfo {volume} {5}},\ \bibinfo {pages} {3400}
  (\bibinfo {year} {2014})}\BibitemShut {NoStop}%
\bibitem [{\citenamefont {Luttinger}(1958)}]{p40}%
  \BibitemOpen
  \bibfield  {author} {\bibinfo {author} {\bibfnamefont {J.~M.}\ \bibnamefont
  {Luttinger}},\ }\href@noop {} {\bibfield  {journal} {\bibinfo  {journal}
  {Phys. Rev.},\ }\textbf {\bibinfo {volume} {112}},\ \bibinfo {pages} {739}
  (\bibinfo {year} {1958})}\BibitemShut {NoStop}%
\bibitem [{\citenamefont {Kanazawa}\ \emph {et~al.}(2011)\citenamefont
  {Kanazawa}, \citenamefont {Onose}, \citenamefont {Arima}, \citenamefont
  {Okuyama}, \citenamefont {Ohoyama}, \citenamefont {Wakimoto}, \citenamefont
  {Kakurai}, \citenamefont {Ishiwata},\ and\ \citenamefont {Tokura}}]{p42}%
  \BibitemOpen
  \bibfield  {author} {\bibinfo {author} {\bibfnamefont {N.}~\bibnamefont
  {Kanazawa}}, \bibinfo {author} {\bibfnamefont {Y.}~\bibnamefont {Onose}},
  \bibinfo {author} {\bibfnamefont {T.}~\bibnamefont {Arima}}, \bibinfo
  {author} {\bibfnamefont {D.}~\bibnamefont {Okuyama}}, \bibinfo {author}
  {\bibfnamefont {K.}~\bibnamefont {Ohoyama}}, \bibinfo {author} {\bibfnamefont
  {S.}~\bibnamefont {Wakimoto}}, \bibinfo {author} {\bibfnamefont
  {K.}~\bibnamefont {Kakurai}}, \bibinfo {author} {\bibfnamefont
  {S.}~\bibnamefont {Ishiwata}}, \ and\ \bibinfo {author} {\bibfnamefont
  {Y.}~\bibnamefont {Tokura}},\ }\href@noop {} {\bibfield  {journal} {\bibinfo
  {journal} {Phys. Rev. Lett.},\ }\textbf {\bibinfo {volume} {106}},\ \bibinfo
  {pages} {156603} (\bibinfo {year} {2011})}\BibitemShut {NoStop}%
\bibitem [{\citenamefont {Huang}\ and\ \citenamefont {Chien}(2012)}]{p44}%
  \BibitemOpen
  \bibfield  {author} {\bibinfo {author} {\bibfnamefont {S.~X.}\ \bibnamefont
  {Huang}}\ and\ \bibinfo {author} {\bibfnamefont {C.~L.}\ \bibnamefont
  {Chien}},\ }\href@noop {} {\bibfield  {journal} {\bibinfo  {journal} {Phys.
  Rev. Lett.},\ }\textbf {\bibinfo {volume} {108}},\ \bibinfo {pages} {267201}
  (\bibinfo {year} {2012})}\BibitemShut {NoStop}%
\bibitem [{\citenamefont {Kumar}\ \emph {et~al.}(2020)\citenamefont {Kumar},
  \citenamefont {Kumar}, \citenamefont {Reehuis}, \citenamefont {Gayles},
  \citenamefont {Sukhanov}, \citenamefont {Hoser}, \citenamefont {Damay},
  \citenamefont {Shekhar}, \citenamefont {Adler},\ and\ \citenamefont
  {Felser}}]{kumar2020detection}%
  \BibitemOpen
  \bibfield  {author} {\bibinfo {author} {\bibfnamefont {V.}~\bibnamefont
  {Kumar}}, \bibinfo {author} {\bibfnamefont {N.}~\bibnamefont {Kumar}},
  \bibinfo {author} {\bibfnamefont {M.}~\bibnamefont {Reehuis}}, \bibinfo
  {author} {\bibfnamefont {J.}~\bibnamefont {Gayles}}, \bibinfo {author}
  {\bibfnamefont {A.}~\bibnamefont {Sukhanov}}, \bibinfo {author}
  {\bibfnamefont {A.}~\bibnamefont {Hoser}}, \bibinfo {author} {\bibfnamefont
  {F.}~\bibnamefont {Damay}}, \bibinfo {author} {\bibfnamefont
  {C.}~\bibnamefont {Shekhar}}, \bibinfo {author} {\bibfnamefont
  {P.}~\bibnamefont {Adler}}, \ and\ \bibinfo {author} {\bibfnamefont
  {C.}~\bibnamefont {Felser}},\ }\href@noop {} {\bibfield  {journal} {\bibinfo
  {journal} {Phys. Rev. B},\ }\textbf {\bibinfo {volume} {101}},\ \bibinfo
  {pages} {014424} (\bibinfo {year} {2020})}\BibitemShut {NoStop}%
\bibitem [{\citenamefont {Suzuki}\ \emph {et~al.}(2016)\citenamefont {Suzuki},
  \citenamefont {Chisnell}, \citenamefont {Devarakonda}, \citenamefont {Liu},
  \citenamefont {Feng}, \citenamefont {Xiao}, \citenamefont {Lynn},\ and\
  \citenamefont {Checkelsky}}]{suzuki2016}%
  \BibitemOpen
  \bibfield  {author} {\bibinfo {author} {\bibfnamefont {T.}~\bibnamefont
  {Suzuki}}, \bibinfo {author} {\bibfnamefont {R.}~\bibnamefont {Chisnell}},
  \bibinfo {author} {\bibfnamefont {A.}~\bibnamefont {Devarakonda}}, \bibinfo
  {author} {\bibfnamefont {Y.-T.}\ \bibnamefont {Liu}}, \bibinfo {author}
  {\bibfnamefont {W.}~\bibnamefont {Feng}}, \bibinfo {author} {\bibfnamefont
  {D.}~\bibnamefont {Xiao}}, \bibinfo {author} {\bibfnamefont {J.~W.}\
  \bibnamefont {Lynn}}, \ and\ \bibinfo {author} {\bibfnamefont
  {J.}~\bibnamefont {Checkelsky}},\ }\href@noop {} {\bibfield  {journal}
  {\bibinfo  {journal} {Nat. Phys.},\ }\textbf {\bibinfo {volume} {12}},\
  \bibinfo {pages} {1119} (\bibinfo {year} {2016})}\BibitemShut {NoStop}%
\bibitem [{\citenamefont {Neupane}\ \emph {et~al.}(2016)\citenamefont
  {Neupane}, \citenamefont {Belopolski}, \citenamefont {Hosen}, \citenamefont
  {Sanchez}, \citenamefont {Sankar}, \citenamefont {Szlawska}, \citenamefont
  {Xu}, \citenamefont {Dimitri}, \citenamefont {Dhakal}, \citenamefont
  {Maldonado} \emph {et~al.}}]{zrsis2016}%
  \BibitemOpen
  \bibfield  {author} {\bibinfo {author} {\bibfnamefont {M.}~\bibnamefont
  {Neupane}}, \bibinfo {author} {\bibfnamefont {I.}~\bibnamefont {Belopolski}},
  \bibinfo {author} {\bibfnamefont {M.~M.}\ \bibnamefont {Hosen}}, \bibinfo
  {author} {\bibfnamefont {D.~S.}\ \bibnamefont {Sanchez}}, \bibinfo {author}
  {\bibfnamefont {R.}~\bibnamefont {Sankar}}, \bibinfo {author} {\bibfnamefont
  {M.}~\bibnamefont {Szlawska}}, \bibinfo {author} {\bibfnamefont {S.-Y.}\
  \bibnamefont {Xu}}, \bibinfo {author} {\bibfnamefont {K.}~\bibnamefont
  {Dimitri}}, \bibinfo {author} {\bibfnamefont {N.}~\bibnamefont {Dhakal}},
  \bibinfo {author} {\bibfnamefont {P.}~\bibnamefont {Maldonado}},  \emph
  {et~al.},\ }\href@noop {} {\bibfield  {journal} {\bibinfo  {journal} {Phys.
  Rev. B},\ }\textbf {\bibinfo {volume} {93}},\ \bibinfo {pages} {201104}
  (\bibinfo {year} {2016})}\BibitemShut {NoStop}%
\bibitem [{\citenamefont {Liu}\ \emph {et~al.}(2018)\citenamefont {Liu},
  \citenamefont {Jin}, \citenamefont {Dai}, \citenamefont {Chen},\ and\
  \citenamefont {Zhang}}]{capd2018}%
  \BibitemOpen
  \bibfield  {author} {\bibinfo {author} {\bibfnamefont {G.}~\bibnamefont
  {Liu}}, \bibinfo {author} {\bibfnamefont {L.}~\bibnamefont {Jin}}, \bibinfo
  {author} {\bibfnamefont {X.}~\bibnamefont {Dai}}, \bibinfo {author}
  {\bibfnamefont {G.}~\bibnamefont {Chen}}, \ and\ \bibinfo {author}
  {\bibfnamefont {X.}~\bibnamefont {Zhang}},\ }\href@noop {} {\bibfield
  {journal} {\bibinfo  {journal} {Phys. Rev. B},\ }\textbf {\bibinfo {volume}
  {98}},\ \bibinfo {pages} {075157} (\bibinfo {year} {2018})}\BibitemShut
  {NoStop}%
\bibitem [{\citenamefont {Kida}\ \emph {et~al.}(2011)\citenamefont {Kida},
  \citenamefont {Fenner}, \citenamefont {Dee}, \citenamefont {Terasaki},
  \citenamefont {Hagiwara},\ and\ \citenamefont {Wills}}]{fe3sn22011}%
  \BibitemOpen
  \bibfield  {author} {\bibinfo {author} {\bibfnamefont {T.}~\bibnamefont
  {Kida}}, \bibinfo {author} {\bibfnamefont {L.}~\bibnamefont {Fenner}},
  \bibinfo {author} {\bibfnamefont {A.}~\bibnamefont {Dee}}, \bibinfo {author}
  {\bibfnamefont {I.}~\bibnamefont {Terasaki}}, \bibinfo {author}
  {\bibfnamefont {M.}~\bibnamefont {Hagiwara}}, \ and\ \bibinfo {author}
  {\bibfnamefont {A.}~\bibnamefont {Wills}},\ }\href@noop {} {\bibfield
  {journal} {\bibinfo  {journal} {J. Phys. Condens. Matter},\ }\textbf
  {\bibinfo {volume} {23}},\ \bibinfo {pages} {112205} (\bibinfo {year}
  {2011})}\BibitemShut {NoStop}%
\bibitem [{\citenamefont {Nakatsuji}\ \emph {et~al.}(2015)\citenamefont
  {Nakatsuji}, \citenamefont {Kiyohara},\ and\ \citenamefont
  {Higo}}]{mn3sn2015}%
  \BibitemOpen
  \bibfield  {author} {\bibinfo {author} {\bibfnamefont {S.}~\bibnamefont
  {Nakatsuji}}, \bibinfo {author} {\bibfnamefont {N.}~\bibnamefont {Kiyohara}},
  \ and\ \bibinfo {author} {\bibfnamefont {T.}~\bibnamefont {Higo}},\
  }\href@noop {} {\bibfield  {journal} {\bibinfo  {journal} {Nat.},\ }\textbf
  {\bibinfo {volume} {527}},\ \bibinfo {pages} {212} (\bibinfo {year}
  {2015})}\BibitemShut {NoStop}%
\bibitem [{\citenamefont {Shao}\ \emph {et~al.}(2021)\citenamefont {Shao},
  \citenamefont {Guo}, \citenamefont {Wu}, \citenamefont {Nie}, \citenamefont
  {Sun}, \citenamefont {Weng},\ and\ \citenamefont {Wang}}]{socgapnacabi}%
  \BibitemOpen
  \bibfield  {author} {\bibinfo {author} {\bibfnamefont {D.}~\bibnamefont
  {Shao}}, \bibinfo {author} {\bibfnamefont {Z.}~\bibnamefont {Guo}}, \bibinfo
  {author} {\bibfnamefont {X.}~\bibnamefont {Wu}}, \bibinfo {author}
  {\bibfnamefont {S.}~\bibnamefont {Nie}}, \bibinfo {author} {\bibfnamefont
  {J.}~\bibnamefont {Sun}}, \bibinfo {author} {\bibfnamefont {H.}~\bibnamefont
  {Weng}}, \ and\ \bibinfo {author} {\bibfnamefont {Z.}~\bibnamefont {Wang}},\
  }\href@noop {} {\bibfield  {journal} {\bibinfo  {journal} {Phys. Rev. Res.},\
  }\textbf {\bibinfo {volume} {3}},\ \bibinfo {pages} {013278} (\bibinfo {year}
  {2021})}\BibitemShut {NoStop}%
\bibitem [{\citenamefont {Noky}\ and\ \citenamefont
  {Sun}(2019)}]{linearresponse2019}%
  \BibitemOpen
  \bibfield  {author} {\bibinfo {author} {\bibfnamefont {J.}~\bibnamefont
  {Noky}}\ and\ \bibinfo {author} {\bibfnamefont {Y.}~\bibnamefont {Sun}},\
  }\href@noop {} {\bibfield  {journal} {\bibinfo  {journal} {Appl. Sci.},\
  }\textbf {\bibinfo {volume} {9}},\ \bibinfo {pages} {4832} (\bibinfo {year}
  {2019})}\BibitemShut {NoStop}%
\bibitem [{\citenamefont {Nielsen}\ and\ \citenamefont
  {Ninomiya}(1981)}]{ninomiya1981}%
  \BibitemOpen
  \bibfield  {author} {\bibinfo {author} {\bibfnamefont {H.~B.}\ \bibnamefont
  {Nielsen}}\ and\ \bibinfo {author} {\bibfnamefont {M.}~\bibnamefont
  {Ninomiya}},\ }\href@noop {} {\bibfield  {journal} {\bibinfo  {journal}
  {Phys. Rev. B},\ }\textbf {\bibinfo {volume} {105}},\ \bibinfo {pages} {219}
  (\bibinfo {year} {1981})}\BibitemShut {NoStop}%
\end{thebibliography}%
\end{document}


\title{\bf{\Large{Supplementary Information}}\\
\smallskip
\smallskip
\smallskip
\smallskip
Unconventional anomalous Hall effect in itinerant hexagonal polar magnet Y$_3$Co$_8$Sn$_4$}

\author{Afsar Ahmed  }
\affiliation{CMP Division, Saha Institute of Nuclear Physics, A CI of Homi Bhabha National Institute, Kolkata 700064, India}
\author{Jyoti Sharma}
\affiliation{Department of Physics, Indian Institute of Technology Bombay, Mumbai 400076, India}
\author{Arnab Bhattacharya }
\affiliation{CMP Division, Saha Institute of Nuclear Physics, A CI of Homi Bhabha National Institute, Kolkata 700064, India}
\author{Anis Biswas}
\affiliation{Ames National Laboratory, Iowa State University, Ames, Iowa 50011, USA}
\author{Tukai Singha}
\affiliation{CMP Division, Saha Institute of Nuclear Physics, A CI of Homi Bhabha National Institute, Kolkata 700064, India}
\author{Yaroslav Mudryk}
\affiliation{Ames National Laboratory, Iowa State University, Ames, Iowa 50011, USA}
\author{Aftab Alam}\email{aftab@iitb.ac.in}
\affiliation{Department of Physics, Indian Institute of Technology Bombay, Mumbai 400076, India}
\author{ I. Das}\email{indranil.das@saha.ac.in}
\affiliation{CMP Division, Saha Institute of Nuclear Physics, A CI of Homi Bhabha National Institute, Kolkata 700064, India}

\begin{abstract}

\end{abstract}

\maketitle
{Here, we present further auxiliary details about the experimental and computational method/procedure, structural analysis, magnetization and magnetotrasnport data, along with the {\it ab-initio} results on the electronic/topological features of Y$_3$Co$_8$Sn$_4$ for structures above and below 53 K. }

{\bf Experimental details -}
Polycrystalline samples of Y$_3$Co$_8$Sn$_4$ were synthesized by conventional arc-melting of high purity 99.9$\%$ constituent elements in Argon (Ar) atmosphere followed by annealing at 800 $^{o}$C for four week in evacuated sealed quartz tube and then quenched into ice-water. The crystallographic structure and the phase purity of the samples were analyzed by X-ray diffraction (XRD) pattern collected at room temperature with Cu-$K\alpha$ radiation using Rigaku TTRX-III diffractometer. The elemental composition and homogeneity of the samples were checked by transmission electron microscope (TEM) equipped with an energy-dispersive X-ray (EDX) spectrometer. DC magnetic properties were characterized using a superconducting quantum interference device (SQUID-VSM) in the temperature (\textit{T}) range of 2 - 300 K and field range of $\pm$ 7 T. The ac magnetic susceptibility measurements were carried out using ac equipped SQUID. Magnetotransport measurements were performed on several samples of dimension, length $\times$ width $\times$ thickness $\approx$ 3.2 - 3.5 mm $\times$ 1.5 - 1.6 mm $\times$ 0.3 - 0.35 mm by a 9 T physical property measurement system (PPMS) using a standard four probe arrangement under an applied dc of 20 $mA$. All the current and voltage probe contacts were taken using Au wires and before measurements, the ohmic nature of the contacts were checked over the entire \textit{T} range. For measuring the field dependency of each isotherms, the samples were cool down from room temperature to the desired temperature in absence of $H$. The transverse resistivity ($\rho_{xy}$) was symmetrized using the relation, $\rho_{xy} = \left[\rho_{xy}(+H) - \rho_{xy}(-H)\right]/2$.

{\bf Computational details -}
To investigate the electronic structure of Y$_3$Co$_8$Sn$_4$ compound in different magnetic configurations, we have performed the ab-initio calculations within the density functional theory (DFT) framework (below and above 53 K ($T_C$)), based on the projector-augmented wave (PAW), using the Vienna Ab initio Simulation Package (VASP). Generalized-gradient approximation by Perdew-Burke-Ernzerhof has been used to describe the exchange and correlation. The lattice parameters (a,b,c) in both the phases below and above ($T_C$) have been optimized, such as for phase below ($T_C$), a=b=8.856 $\AA$, c=7.458 $\AA$, whereas for phase above ($T_C$), a=b=8.878 $\AA$  and c=7.486 $\AA$,  respectively. Brillouin zone (BZ) integration has been performed on a 6×6×8 $\Gamma$-centered k-mesh with the total energy convergence up to $10^{-6}$ eV. A plane wave energy cutoff of 410 eV was used for all the calculations. A tight-binding Hamiltonian has been constructed using the pre-converged DFT results, using the Wannier90 package. Maximally localized Wannier functions for Co(d), Y(d) and Sn(p) have been used as the basis of tight-binding Hamiltonian. WANNIERTOOLS package has been employed to simulate the topological properties (like nodal points, Weyl chirality, bulk band dispersion etc.), and Anomalous Hall conductivity (AHC). AHC has been calculated as the sum of Berry curvature over all the occupied bands, with 201×201×201 k-point grid.

\begin{figure*}[h]
\begin{center}
\includegraphics[width=.95\textwidth]{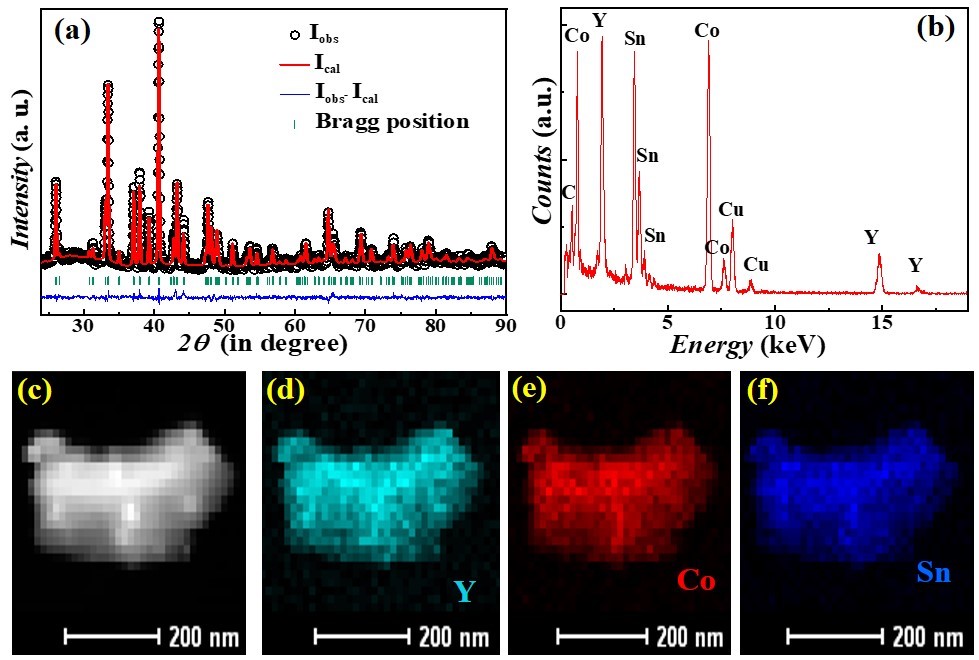}
\end{center}
\caption[]
{\label{SF1}For Y$_3$Co$_8$Sn$_4$, (a) Room temperature XRD pattern along with the Rietveld refinement data. (b) EDX spectra.(c) STEM-HAADF image of the particle on which the elemental mapping was taken. Elemental profile of (d) Y (e) Co and (f) Sn. }
\end{figure*}

\begin{figure*}[h]
\begin{center}
\includegraphics[width=.85\textwidth]{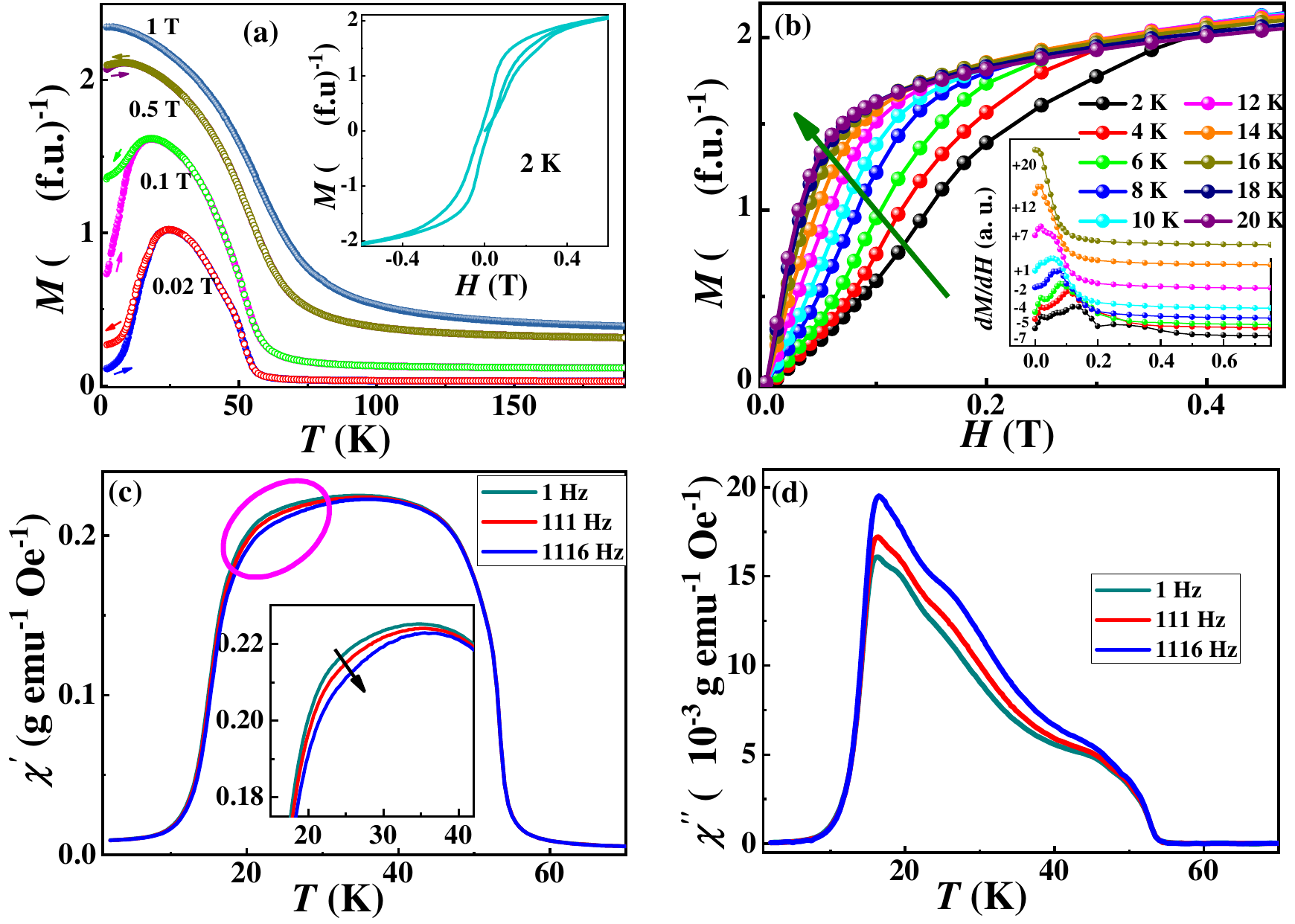}
\end{center}
\caption[]
{\label{SF3}For Y$_3$Co$_8$Sn$_4$ (a) $M$ vs. $T$ under different $H$ in ZFC and FC protocols. Inset shows enlarged view of 2 K $M$-$H$. (b) Isothermal $M$-$H$ curves in the low temperature range (2- 20 K, $\Delta T$ =2 K). Inset shows $dM/dH$ vs. $H$ for in the same temperature range. It is to be noted that an offset is created along y-axis for better visualization. (c) Temperature dependence of the real part of ac susceptibility ($\chi'(T)$) at $f$ = 1, 111, 1116 Hz. Inset shows frequency dependent peak shift of $\chi'(T)$. (d) Temperature dependence of the imaginary component of ac susceptibility $\chi^{\prime\prime}(T)$ at  $f$ = 1, 111, 1116 Hz. }
\end{figure*}

\begin{table*}[h]
\begin{center}
\caption[]
{\label{ST1} List of crystallographic parameters after Rietveld refinement of the XRD data of  Y$_3$Co$_8$Sn$_4$ at room temperature. The numbers in parenthesis are the uncertainty in the refinement. }
\begin{tabular}{p{5.0 cm}p{5.5 cm}p{5.5 cm}}
\hline
\hline
\noalign{\smallskip}
Space Group & P6$_3$mc (No. 186)\\
Lattice parameters: & a = b = 8.875(1) \r{A} \\
 & c = 7.474(2) \r{A} \\ \\
$R_{f} = $& $ 4.08$ \\
$R_{Bragg} =$ & $7.17$ \\
$\chi^2 =$  & 1.81 \\ \\
\noalign{\smallskip}
\noalign{\smallskip}
&\textbf{\indent \indent Atomic Coordinates}&\\
\end{tabular}
\begin{tabular}{p{2.65 cm}p{2.65 cm}p{2.65 cm}p{2.65 cm}p{2.65 cm}p{2.65 cm}}
Atom  &Wyckoff Position &  x   &   y  &   z  & occupancy\\
\hline
\noalign{\smallskip}
Y & 6c & 0.471(7) &  0.529(3) &  0.000 & 1\\
Co1& 6c & 0.817(8) &  0.182(2) &  0.833(9) & 1 \\
Co2& 6c & 0.957(6) &  0.042(4) &  0.508(1) & 1 \\         
Co3& 2b & 0.333(3) &  0.666(7) &   0.646(7) & 1 \\        
Co4& 2a & 0.000 &  0.000 &  0.312(4) & 1  \\ 
Sn1& 6c & 0.812(4) & 0.189(6) &  0.793(2) & 1  \\         
Sn2& 2b & 0.333(3) &  0.666(7) &   0.289(5) & 1 \\
\noalign{\smallskip}
\noalign{\smallskip}
\hline
\hline
\end{tabular}
\end{center}
\end{table*}

{\bf Structure -} The room temperature X-ray diffraction data was analyzed by Rietveld refinement using FullProf software as displayed in Fig.\ref{SF1}(a). Y$_3$Co$_8$Sn$_4$ crystallizes in a polar hexagonal structure, similar to the Lu$_3$Co$_{7.77}$Sn$_4$ type, with all the peaks indexed considering the $P6_3mc$ space group. The lattice parameters and Wyckoff positions are enlisted in Table \ref{ST1}, and are aligned well with previous studies \cite{canepa2000crystal,takagi2018multiple,canepa2004ferromagnetic}. Y$_3$Co$_8$Sn$_4$ is a distorted version of the BaLi$_4$ structure (space group: $P6_3/mmc$), influenced by the arrangement of Y and Sn atoms. The Y atoms occupy the Ba site (6$c$ Wyckoff position), while the 4$f$ Wyckoff position (Li$_2$ site) is split into two 2$b$ sites, which are filled by Sn$_2$ and Co$_3$. Furthermore, the 12$d$ Wyckoff position in $P6_3/mmc$ splits into 6$c$, where Sn$_1$ and Co$_2$ are located. This splitting of atomic positions causes the centrosymmetric $P6_3/mmc$ space group to transform into the non-centrosymmetric $P6_3mc$, due to internal distortions.

 Compositional analysis was conducted on different regions of the sample, and one of the EDX spectra is presented in Fig. \ref{SF1}(b). From the EDX analysis the composition is found to be Y$_{2.97}$Co$_{8.13}$Sn$_{4.07}$, which confirms uniformity within 2 \% of the stoichiometric ratio. Figure \ref{SF1}(c) shows the scanning TEM high-angle annular dark-field (STEM-HAADF) image of the particle, on which elemental mapping was performed. Elemental maps of Y, Co, and Sn confirm the compound's homogeneity, as shown in Fig. \ref{SF1}(d)-(f).

{\bf Magnetization -}
The variation of magnetization ($M$) with temperature ($T$) for various magnetic fields ($H$ = 0.02, 0.1, 0.5, and 1 T) is measured under zero-field cooling (ZFC) and field cooling (FC) protocols, as shown in Fig. \ref{SF3}(a). Irreversibility between the ZFC and FC measurements is observed up to 0.5 T, with the peak of each thermomagnetic curve shifting to lower temperatures indicating an antiferromagnetic (AFM) correlation. Up to $H = 0.1$ T, the linear increase in the virgin curve of the 2 K isothermal $M$-$H$ data reflects the AFM behavior. However, the presence of a valley before reaching saturation (field-induced ferromagnetism) suggests the coexistence of another magnetic phase. Figure \ref{SF3}(b) shows isothermal magnetization curves at low temperatures (2–20 K, $\Delta T$ = 2 K), highlighting the AFM to field-induced FM transition, consistent with earlier findings \cite{canepa2004ferromagnetic}, with the valley becoming suppressed.  As the temperature rises, the isotherms shift towards lower $H$ values, indicating a reduced field requirement to achieve the field-induced FM state. The multiple magnetic orderings in the low-temperature region are clearly illustrated in the $dM/dH$ vs. $H$ plot shown in the inset of Fig. \ref{SF3}(b). At 2 K, a prominent peak with two shoulder-like features is observed, which shifts to lower field values at higher temperatures and nearly vanishes by 16 K. This suggests modifications of non-trivial magnetic states with changes in both field and temperature, consistent with the SANS report by Takagi et al.\cite{takagi2018multiple}.

To gain deeper insight into the modifications of magnetic ordering, we performed temperature and field-dependent ac-susceptibility measurements. Figure \ref{SF3}(c) and \ref{SF3}(d) display the temperature-dependent real and imaginary components of ac susceptibility, respectively, for frequencies of f = 1, 111, and 1116 Hz. The real part, $\chi^{\prime}$ rises sharply at 53 K, marking the onset of ferromagnetic ordering, and remains nearly constant between 25 K and 45 K. Below 25 K, $\chi^{\prime}$ decreases, exhibiting a frequency-dependent behaviour (inset of Fig. \ref{SF3}(c)), with a change in slope observed around 15 K. The frequency dependence of $\chi^{\prime}$ may stem from the frustration of short-range ferromagnetic interactions embedded in a long-range antiferromagnetic ordered state, ruling out the spin-glass nature of the compound \cite{canepa2004ferromagnetic}. Interestingly, besides the peak at 16 K, multiple cusps between 16 - 53 K are observed in $\chi^{\prime\prime}$, potentially arising from various magnetic orderings.

\begin{figure*}[h]
\begin{center}
\includegraphics[width=.95\textwidth]{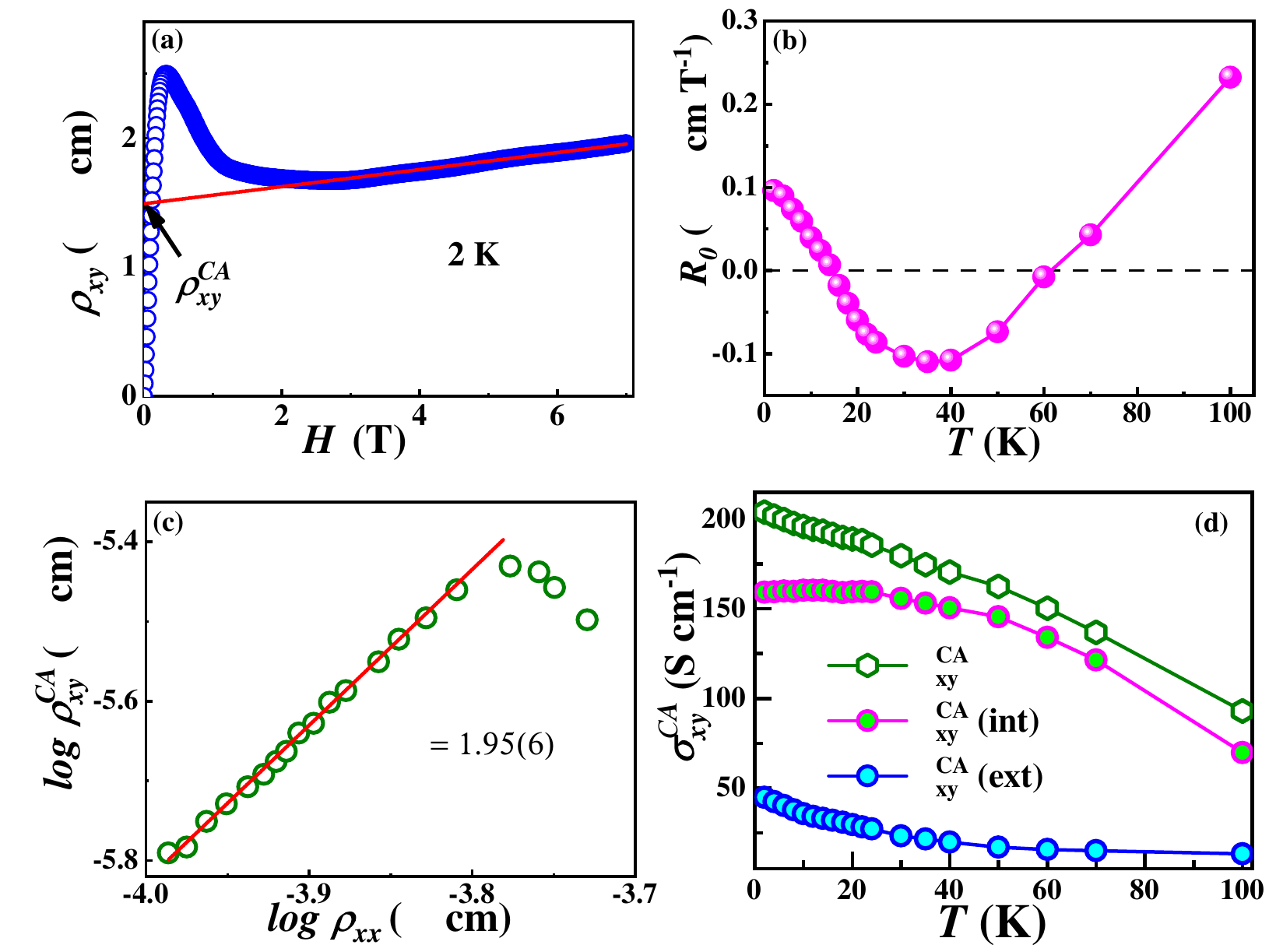}
\end{center}
\caption[]
{\label{SF4}For Y$_3$Co$_8$Sn$_4$,  (a) Hall resistivity vs. $H$ at 2 K (blue circle). Red line indicates a linear fitting of $\rho_{xy}$ in the high field range ( 5 - 7 T). (b) Variation of normal Hall coefficient ($R_0$) with $T$. (c) Power law fitting of $\rho_{xy}^{CA}$ with $\rho_{xx}$ in a log-log scale. (d) The intrinsic and extrinsic contribution to the total AHC. }
\end{figure*}

{\bf Magnetotransport -} Figure \ref{SF4}(a) depicts the transverse resistivity $\rho_{xy}$ as a function of $H$ for 2 K. The linear increase in $\rho_{xy}$ above H = 3 T, is attributable to the Lorentz force acting on charge carriers. Interestingly, unlike the isothermal $M(H)$ curve, a distinct bump is observed below 2.5 T, indicating a characteristic feature of topological materials\cite{xu2021unconventional,p34,p35}. Given the absence of non-trivial magnetic ordering in the compound at higher $H$, we fitted $\rho_{xy}$ from 3 to 7 T, with the intercept yielding the conventional anomalous Hall resistivity ($\rho_{xy}^{CA}$) and the slope corresponding to the normal Hall coefficient $R_0$. We collected $R_0$ for each Hall isotherm following the same procedure and plotted the results in Fig. \ref{SF4}(b), where a crossover in the dominant carrier types is observed around 15 K and 60 K. Figure \ref{SF4}(c) shows the log($\rho^A_{xy}$) vs. log($\rho_{xx}$) curve, fitted using the power-law relation $\rho_{xy} \propto \rho_{xx}^\alpha$, which yields $\alpha \approx 1.95(6)$. This indicates a predominant contribution arising from a combination of intrinsic and/or side-jump mechanisms in the anomalous Hall effect (AHE) \cite{p13}. To further distinguish the contributions to the anomalous Hall conductivity (AHC), we employed the \textit{TYJ} scaling relation, expressed as $\sigma_{xy}^{A} = -a\sigma_{xx0}^{-1}\sigma_{xx}^2 - b$. Here, $\sigma_{xx0}$ denotes the residual conductivity, while $b$ corresponds to the combined contributions from the intrinsic ($\sigma_{int}$) and side-jump ($\sigma_{sj}$) mechanisms \cite{p13, p14, p12, p17}. Using this scaling approach, the parameters $a$ and $b$ were determined. Subsequently, the individual contributions of the side-jump and intrinsic mechanisms were calculated using the relations: $\sigma_{sk} = \frac{a\rho_{xx0}}{(\rho_{xy}^{CA})^2 + \rho_{xx}^2}$ and $\sigma_{int} = \frac{b\rho_{xx}^2}{(\rho_{xy}^{CA})^2 + \rho_{xx}^2}$ as detailed in \cite{chakraborty2022berry}. Figure \ref{SF4}(d) illustrates the temperature dependence of $\sigma_{int}^{CA}$ and $\sigma_{sk}^{CA}$, showing that $\sigma_{int}^{CA}$ reaches a value of 160 S cm$^{-1}$ at 2 K and remains the dominant contribution over the entire temperature range. To quantify the relative contributions of $\sigma_{int}$ and $\sigma_{sj}$, the order of magnitude of $\sigma_{sj}$ was estimated using the relation: ($e^2/hv^{1/3})\epsilon_{SOC}/E_F$, where $\epsilon_{SOC}$ is the spin-orbit coupling energy, and $v$ is the unit cell volume. Using the lattice parameters, the resonant AHC, $e^2/hv^{1/3}$, was calculated as 462 S cm$^{-1}$ \cite{miyasato2007crossover,onoda2006intrinsic}. Assuming $\epsilon_{SOC} = 0.01$ for itinerant magnetic metals, $\sigma_{sj}^A$ was estimated to be approximately 4.62 S cm$^{-1}$, which is two orders of magnitude smaller than $\sigma_{int}^A$. This confirms that the intrinsic contribution predominantly governs the AHE.

\begin{figure*}[h]
\begin{center}
\includegraphics[width=0.7\textwidth]{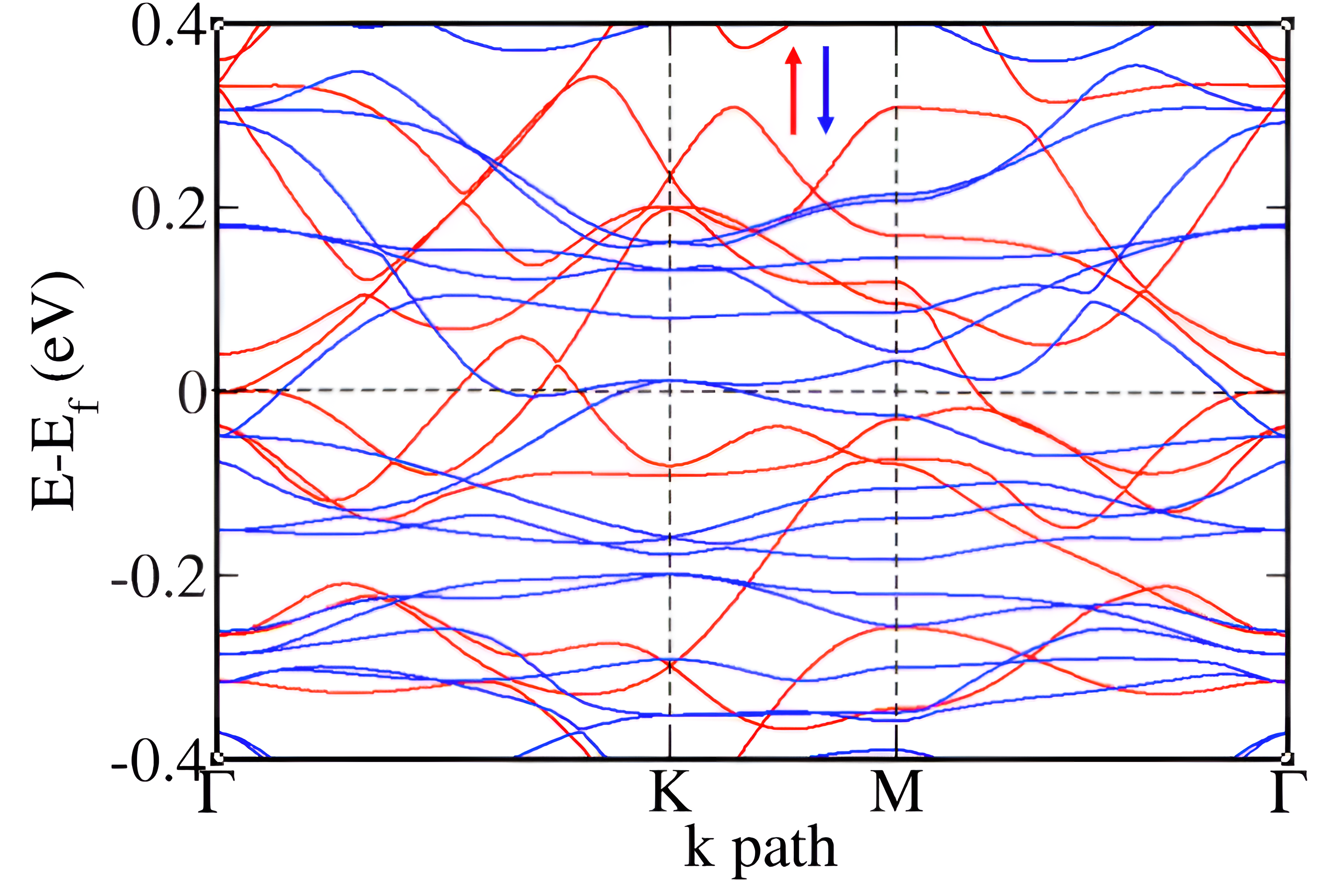}
\end{center}
\caption[]
{\label{Fig.S5}
Electronic bulk band structure without SOC, for the lowest energy planar FiM2 magnetic configuration (below 53 K)  of Y$_3$Co$_8$Sn$_4$. Here red (blue) arrow shows the band for spin up (down) channels.}
\end{figure*}

{\bf Electronic/Topological Features for structure below 53 K:} 
 To elucidate the electronic structure of the present compound below 53 K, we have simulated various magnetic configurations such as non magnetic (NM), ferromagnetic (FM), axial and planar antiferromagnetic and ferrimagnetic phases using the experimental structure\cite{canepa2004ferromagnetic}. Figure \ref{Fig.S5} shows the schematic of these magnetic configurations in the hexagonal unit cell. For better visualization, sublattices involving Co atoms only (i.e. Co1, Co2, Co3, Co4)  are shown, whereas Y and Sn atoms sublattices are omitted, as magnetism is mainly governed by the Co atoms in this compound. The unit cell contains 6 Co1, 6 Co2, 2 Co3 and 2 Co4 atoms comprising of a total of 30 atom (2 formula unit). For axial cases, the moments are aligned out of plane (i.e. (0,0,$m_z$)) while for planar cases, they are aligned in-plane (i.e. ($m_x$,$m_y$,0)). In the AFM case (both axial and planar), 50\% of each Co atoms have spin up while the other 50\% has spin down, yielding a zero net magnetization. In case of FiM1, all the Co1, Co2 and Co4 atoms are aligned along one direction (ferromagnetically coupled) while all Co3 are aligned in the opposite direction with respect to Co1, Co2 and Co4. In case of FiM2, all the Co1 and Co2 are aligned in one direction while Co3 and Co4 atoms are aligned in the opposite direction. Table I of main manuscript displays the simulated relative energies, total and Co projected magnetic moments at different sublattices, and moment per Co atom for the above mentioned magnetic configurations. 
 It should be noted here that, even if we start with a FM configuration with all the moments  aligned in the same direction for all the Co sublattices, it always converges to a either FiM1 or FiM2 configurations, confirming the robustness of the later.  FiM2 (planar) ordering (with Co moments oriented in ab plane (easy plane)) is found to be energetically the most stable configuration. In FiM2 configuration, Co3 (2b site) atoms are antiferromagnetically coupled with respect to the other Co-atoms, which gives rise to a net magnetization of around 3.3 $\mu{B}$/f.u. and average moment per Co to be 0.4 $\mu_B$, which are in good agreement with our experimental results. 
 
\begin{figure*}[h]
\begin{center}
\includegraphics[width=0.8\textwidth]{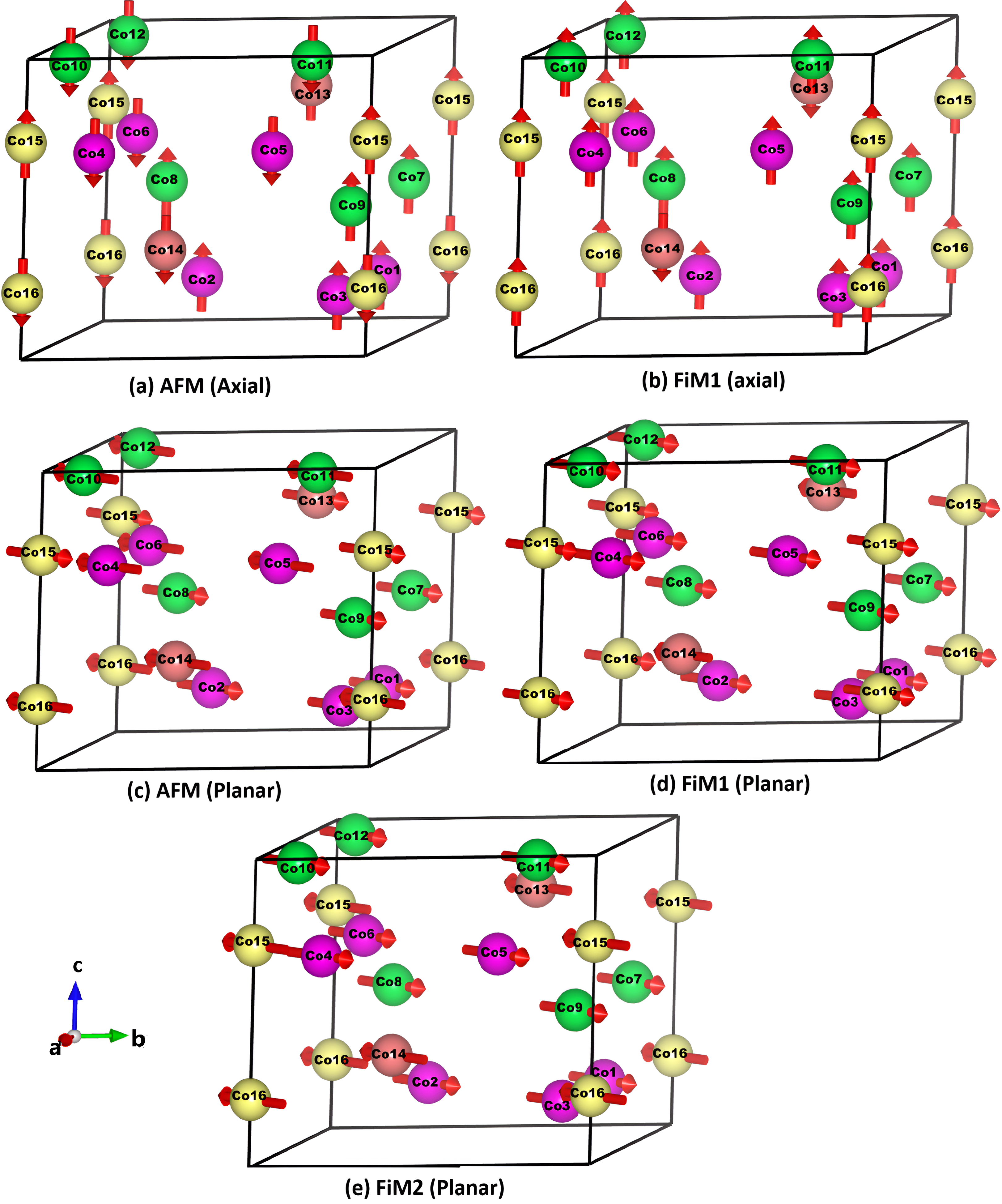}
\end{center}
\caption[]
{\label{Fig.S4}
Various magnetic orderings of Y$_3$Co$_8$Sn$_4$ in a hexagonal unit cell (a) AFM (Axial) (b) FiM1 (Axial) (c) AFM (Planar) (d) FiM1 (Planar) and (e) FiM2 (Planar). For better visualization, only Co atoms in different sublattices (i.e. Co1, Co2, Co3, Co4) are shown, while Y and Sn atoms are omitted. There are 6 Co1, 6 Co2, 2 Co3 and 2 Co4 atoms comprising of a total of 16 Co-atoms in the simulated unit cell (2 formula unit).}
\end{figure*}

Figure \ref{Fig.S5} shows the electronic bulk band structure of Y$_3$Co$_8$Sn$_4$ in the most stable FiM2 configuration without spin-orbit coupling (SOC).

{\bf Electronic/Topological Features for structure above 53 K:} 
To investigate the unconventional transport properties and their mechanism above 53 K i.e. the non/paramagnetic phase, we have simulated the electronic band structure of this phase as well. The electronic bulk band structure of the present compound without SOC in this phase is shown in Fig. S6. It clearly shows a metallic behavior with several linear crossings at/around the Fermi level, which are to be checked further for their trivial or nontrivial topological nature. 

\begin{figure*}[h]
\begin{center}
\includegraphics[width=0.8\textwidth]{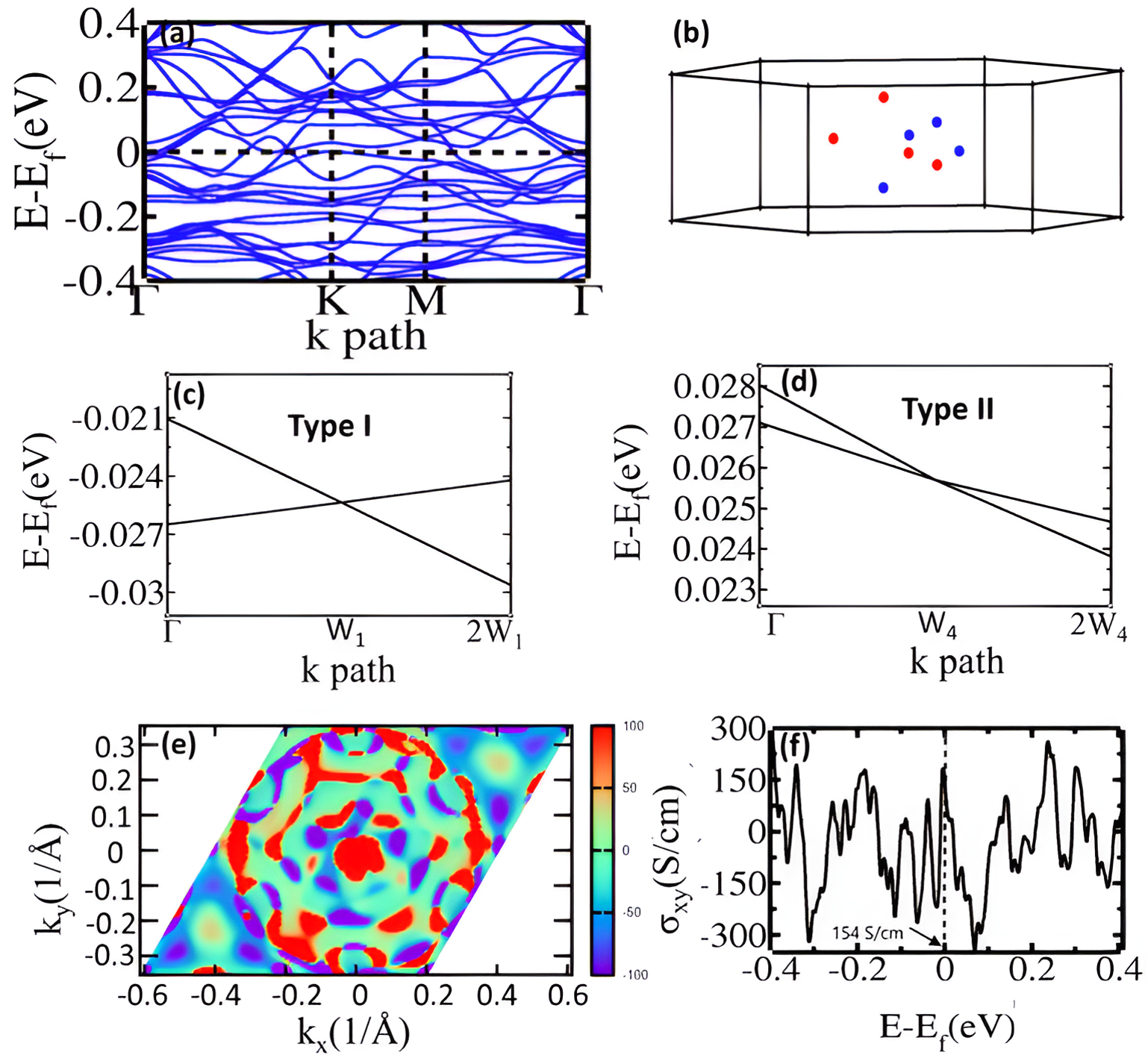}
\end{center}
\caption[]
{\label{Fig.S7}For the nonmagnetic phase of $\mathrm{Y}_3\mathrm{Co}_8\mathrm{Sn}_4$, (a) Electronic bulk band structure including the effect of SOC (b) Distribution of 4 pairs of WPs in the bulk BZ. WPs with +1(-1) chirality are shown by red (blue) dots. (c,d) Bulk band dispersion in the vicinity of a representative type-I and type-II WPs. (e) Berry curvature in the $k_x$-$k_y$ plane. (f) Energy dependence of anomalous Hall conductivity ($\sigma_{xy}$).}
\end{figure*}

Figure \ref{Fig.S7}(a) shows the band structure of Y$_3$Co$_8$Sn$_4$ including SOC effect. A search for topological nodal points in the entire bulk BZ is carried out, which revealed a total of 4 pairs of WPs with opposite chirality (see Fig. S7(b)). Out of these 4 pairs, 1 pair is of type-I, whereas the other 3 are of type-II nature. A zoomed-in view of the band structure in the vicinity of a representative type I and type II WPs are shown in Fig. S7(c,d). Total topological charge associated with these 8 WPs sum up to zero. Table ST2 shows the reciprocal space coordinates, type, energy position (relative to Fermi level (E$_F$)) and chirality of these 4 pairs of WPs above 53 K. One can see from the table that most of these WPs lie close to E$_F$, which is expected to yield large Berry curvature, as shown in Fig. S7(e). Using this Berry curvature, we simulated the anomalous Hall conductivity ($\sigma_{xy}^A$) of Y$_3$Co$_8$Sn$_4$ in the NM phase. This is shown in Fig. S7(f). The simulated value of $\sigma_{xy}^A$ at E$_F$ is found to be 154 S/cm, which agrees fairly well with the measured value (136 S cm$^{-1}$ at 60 K).

\begin{figure*}[h]
\begin{center}
\includegraphics[width=0.7\textwidth]{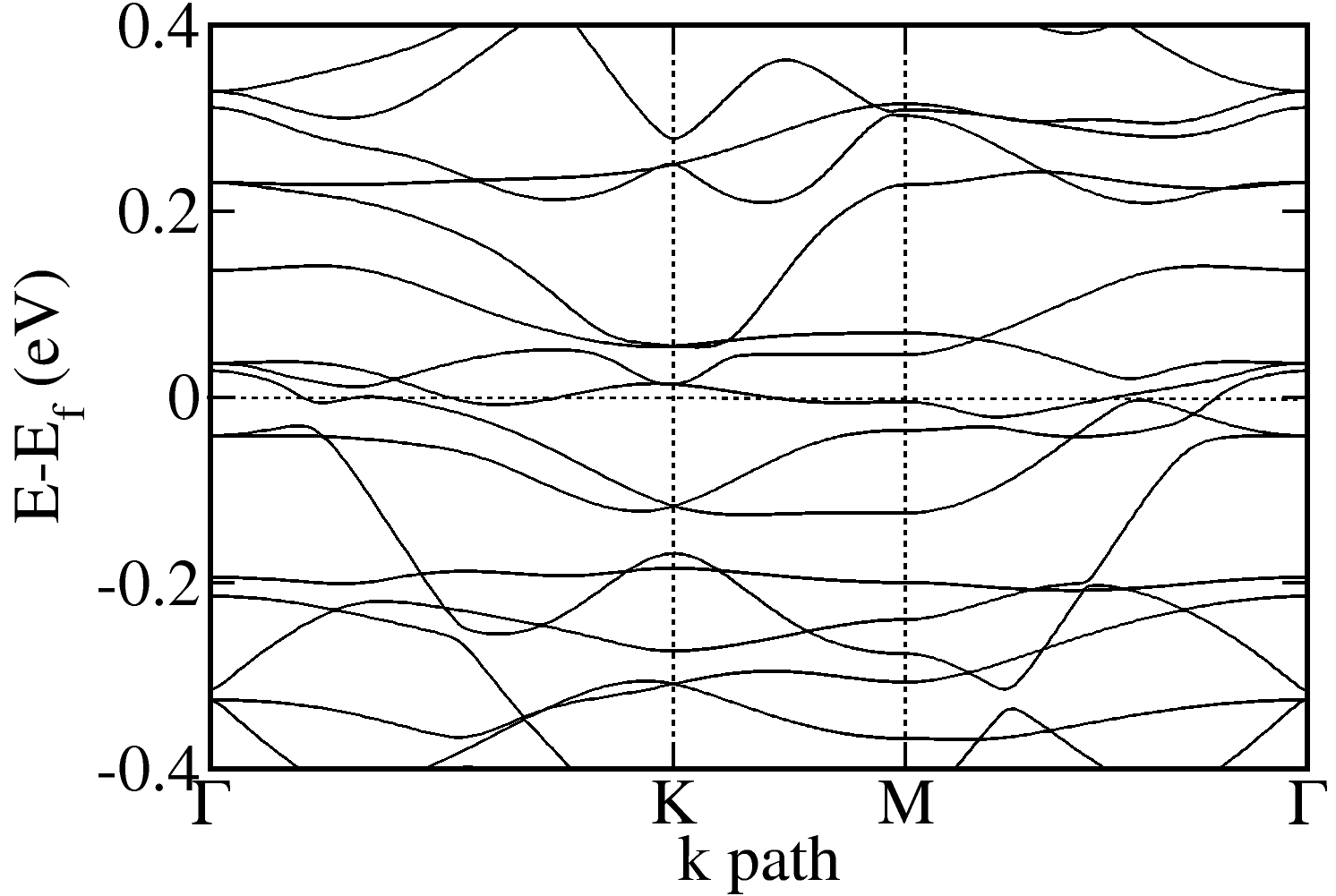}
\end{center}
\caption[]
{\label{Fig.S6}
Electronic bulk band structure without SOC, for the  nonmagnetic phase of Y$_3$Co$_8$Sn$_4$ (above 53 K). }
\end{figure*}

\begin{table*}[t]
	\centering
\caption{For nonmagnetic phase of Y$_3$Co$_8$Sn$_4$, reciprocal space coordinates, type, energy position (relative to E$_F$) and chirality of the four pairs of Weyl points (WPs). }
\label{tableST3}
\begin{ruledtabular}
    	\begin{tabular}{c  c c c c c c}
			WPs &  & Cartesian coordinates of WPs & & Type of WP  & Relative position w.r.to E$_F$ & Chirality   \\
            
            & kx(\AA$^{-1}$) & ky(\AA$^{-1}$) & kz(\AA$^{-1}$) & & (eV) &   \\
            \hline	
            $W_{1}^{-}$ & 0.294	&0.281	&-0.04 & I & -0.023 & -1 \\
           $W_{1}^{+}$ & -0.296	&-0.281	& 0.05 & I & -0.025 & +1 \\
           $W_{2}^{-}$ &  -0.013 &	0.329  &	0.193 & II & 0.013 & -1 \\
           $W_{2}^{+}$ & -0.012 &	0.331 & 	-0.190 & II & 0.012 & +1 \\
           $W_{3}^{-}$ &  -0.071	& 0.154 &	0.082 & II & 0.024 & -1 \\
           $W_{3}^{+}$ &   -0.072 &	0.152 &	-0.080 & II & 0.026 & +1 \\
           $W_{4}^{-}$ & -0.350 &	0.160 &	-0.408 & II & 0.028 & -1 \\
           $W_{4}^{+}$ & -0.348	& 0.163	& 0.405 & II & 0.026 & +1 \\

              \end{tabular}
\end{ruledtabular}
\end{table*}

\bibliography{ref}